%% file: sample63.tex
\documentclass[twocolumn,twocolappendix]{aastex63}
\usepackage{multirow,booktabs}
\usepackage{xcolor}
\accepted{Nov 5, 2021}

\submitjournal{ApJ}
\shorttitle{Spots on young stars}
\shortauthors{Flores et al.}
\graphicspath{{./}{figures/}}

\begin{document}

\title{The Effects of Starspots on Spectroscopic Mass Estimates of Low-mass Young Stars}

\email{caflores@hawaii.edu}
\correspondingauthor{Christian Flores}

\author[0000-0002-8591-472X]{C. Flores}
\author[0000-0002-8293-1428]{M. S. Connelley}
\author[0000-0001-8174-1932]{B. Reipurth}

\affiliation{Institute for Astronomy, University of Hawaii at Manoa, 640 N. Aohoku Place, Hilo, HI 96720, USA \\}

\author[0000-0002-5092-6464]{G. Duch\^ ene}
\affiliation{Astronomy Department, University of California, Berkeley, CA 94720, USA}
\affiliation{Universit\'e Grenoble Alpes, CNRS, Institut de Plan\'etologie et d'Astrophysique, IPAG, 38000 Grenoble, France}



\begin{abstract}
Magnetic fields and mass accretion processes create dark and bright spots on the surface of young stars. These spots manifest as surface thermal inhomogeneities, which alter the global temperature measured on the stars. To understand the effects and implications of these starspots, we conducted a large iSHELL high-resolution infrared spectroscopic survey of T Tauri stars in  Taurus-Auriga and Ophiuchus star-forming regions. From the \textit{K} band spectra, we measured stellar temperatures and magnetic field strengths using a magnetic radiative transfer code. We compared our infrared-derived parameters against literature optical temperatures and found a) a systematic temperature difference between optical and infrared observations, and b) a positive correlation between the magnetic field strengths and the temperature differences. The discrepant temperature measurements imply significant differences in the inferred stellar masses from stellar evolutionary models.  To discern which temperature better predicts the mass of the star, we compared our model-derived masses against dynamical masses measured from ALMA and PdBI for a sub-sample of our sources. From this comparison we conclude that, in the range of stellar masses from 0.3 to 1.3 $\rm M_\odot$, neither infrared nor optical temperatures perfectly reproduce the stellar dynamical masses. But, on average, infrared temperatures produce more precise and accurate stellar masses than optical ones.
\end{abstract}



\section{Introduction} \label{sec:intro}

Young stars have magnetic fields that directly and indirectly produce hot and cold spots on their stellar surfaces. Hot spots produced through accretion appear when circumstellar material channeled from the disk impacts the stellar surface at almost free-fall speed \citep{Gullbring2000,Hartmann2016}. Hot spots can also be produced in the form of stellar plages and faculae due to an increase in surface magnetic activity, similar to what is observed in the Sun \citep{Strassmeier2009}. Meanwhile cold stellar spots appear when magnetic fields inhibit the transport of energy through convection, producing localized dark regions \citep{Berdyugina2005}. 

Although there has been substantial research on the photometric effects of spots in stars, spectroscopic signatures have often been neglected, in part because spectroscopic monitoring of stars is observationally expensive but also because most evolved stars host weak magnetic fields. Indeed, in a study of low-mass stars with ages from less than one Myr to over a Gyr old, \cite{Vidotto2014} found that the average magnetic field strength derived from polarimetry in young stars is thousands of times stronger than the magnetic field value of main-sequence stars. This field difference means that the size and/or temperature contrast of cold spots on young stars, could be considerably larger than what is observed on their more evolved counterparts. Hints of this phenomenon have emerged from individual sources, where significant differences were reported for temperature measurements performed at different wavelengths ($\rm \Delta \, T > 300 $~K) \citep[e.g.,][]{Vacca2011,Gully-Santiago2017,Flores2019,Flores2020}. 

One of the main problems of these discrepant temperature measurements is that the effective temperature of such stars can no longer be obtained from a single small-range spectroscopic observation \citep{Gully-Santiago2017}. For low-mass young stars, this additionally implies that their calculated masses could be systematically offset, as their effective temperatures strongly correlates with their stellar masses. This occurs because low-mass stars contract almost isothermally during the first 5-20 Myrs of their evolution as they descend the Hayashi track. Therefore, understanding and correcting for this effect is of paramount importance as direct mass measurements of young stars are often hard and expensive to obtain and, most of the time, simply impossible \citep{Guilloteau2014,Simon2017,Simon2019,Braun2021}. 

In this paper, we present iSHELL \textit{K} band observations of 40 bright ($\rm K<11$) young stellar sources from Ophiuchus and Taurus-Auriga star forming regions. We mostly selected sources from \cite{Simon2019} and the ODISEA project \citep{Cieza2019} and did our best to try to avoid spectroscopic binaries. The literature optical spectral types for the sources in our sample range from K2 to M4.5. Optical spectral types for the sources in Taurus-Auriga were collected from \cite{White2004,Hartigan2003,Luhman2010,Herczeg2014} while optical measurements for the Ophiuchus sources were obtained from \cite{Bouvier1992,Wilking2005,Torres2006,Erickson2011}. Two stars do not have optical spectral types (GSS 26 and GSS 39), and another source (WSB 82) has only a broad classification of mid-K-type \citep{Esplin2018}. All but one of the young stars in our sample are Class~II sources, the exception is Haro 6-13, which is a Class~I source. 

Our spectroscopic observations with iSHELL are presented in Section \ref{sec:observations}. We derive stellar parameters for the stars using the radiative transfer code MoogStokes \citep{Deen2013} in Section \ref{sec:stellar_params}. In Section \ref{sec:Results}, we combine optical and infrared temperatures with the magnetic stellar evolutionary models of \cite{Feiden2016} to derive stellar masses. The model-derived masses are then compared to independently measured stellar masses from interferometric observations. We discuss our results in Section \ref{sec:discussion} and summarize our findings in Section \ref{sec:conclusion}.
\\

\section{Observations and Data Reduction} \label{sec:observations}


We observed 40 young stars between 2017 October 13 and 2020 October 23 using the high-resolution near-infrared echelle spectrograph iSHELL on IRTF \citep{Rayner2016}. The observations were performed in the $K2$ mode, thus from 2.09 to 2.38 $\mu$m; using the 0\farcs75 slit to achieve a spectral resolution of R $\sim$ $50,000$. 
Due to a known fringing issue in the iSHELL spectra, we decided to adopt the following observing strategy. First, we guided on a young star and integrated on it long enough to obtain a spectrum with $S/N>50$. We stopped guiding and immediately acquired a set of calibration spectra consisting of $K$~band flats and arc-lamp spectra. Then, we observed a nearby telluric A0 standard star within 0.1 airmasses of the young source and integrated on it long enough to achieve a $S/N>100$ spectrum. Afterward, we stopped the guider and obtained a second set of $K$~band calibration files to correct the telluric standard star. The same process was repeated for each star in the sample.

The SNR obtained for the young sources ranges between 56 and 175 with a median value of 105, as reported by \texttt{Spextool v5.0.2} \citep{Cushing2004}. The data were reduced in the same manner as reported in \cite{Flores2020}.

\begin{figure*}[ht!]
\epsscale{1.1}
\plotone{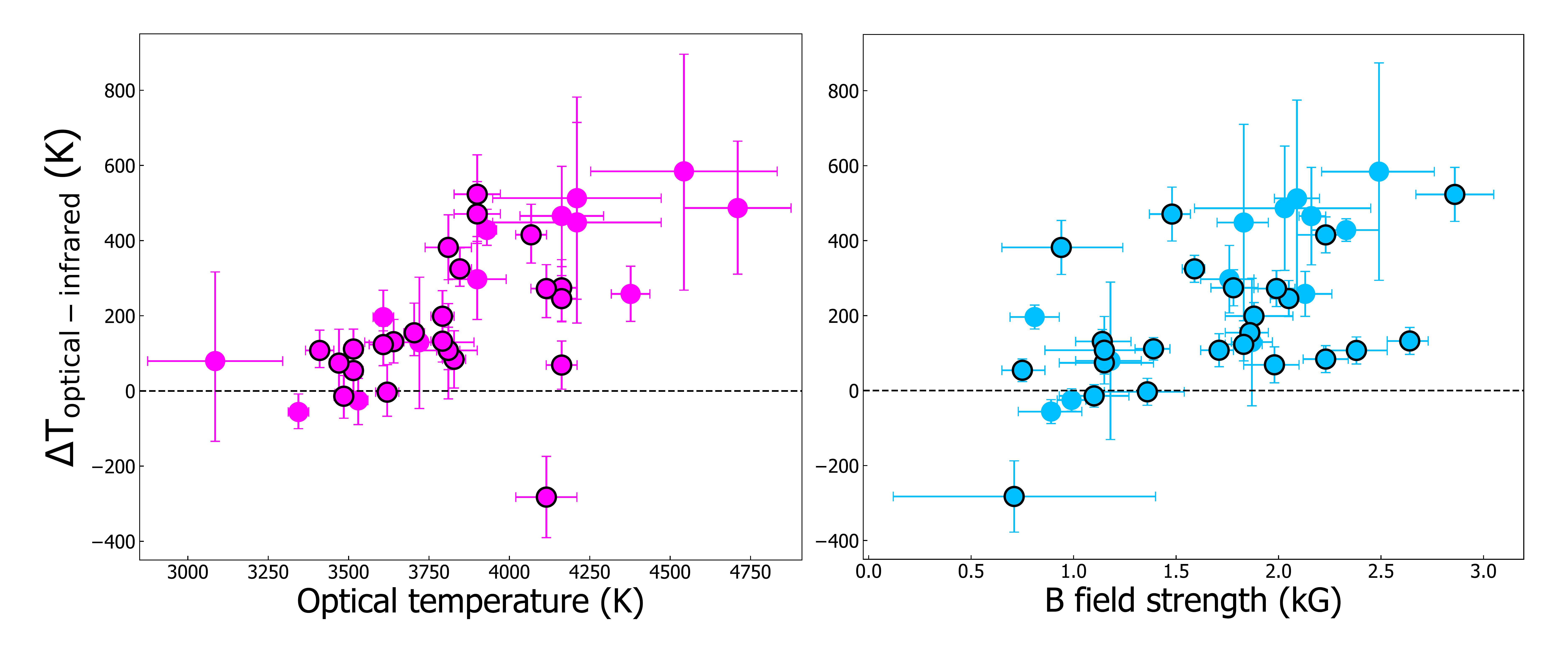}
\caption{Left panel: temperature difference between optical and \textit{K} band measurements as a function of the optical temperature. Right panel: temperature difference between optical and \textit{K} band measurements as a function the magnetic field strengths. Both relations are significantly correlated and show a positive trend. Data points with thick black contours correspond to the sample of 24 stars with measured dynamical masses presented in section \ref{sec:tempr_masses_vs_dynamical}. \label{fig:temperature_difference_optical}}
\end{figure*}

\section{Stellar Parameter Measurements}
\label{sec:stellar_params}
We derive stellar parameters for the 40 young stars using a synthetic stellar spectrum technique that combines MARCS stellar atmospheric models \citep{Gustafsson2008}, the magnetic radiative transfer code MoogStokes \citep{Deen2013}, along with atomic and molecular lines from the VALD3 \citep{Ryabchikova2015,Pakhomov2019} and HITEMP databases \citep{Rothman2010}. 

More details of the modeling technique can be found in \cite{Flores2019,Flores2020} but in brief, we took a two-step approach to fit the observations. First, we varied seven stellar parameters: temperature, gravity, magnetic field strength, projected rotational velocity, \textit{K} band veiling, micro-turbulence, and CO abundance. We performed the fitting using eight wavelength regions, one of which contains the CO overtone starting at 2.294 $\micron$ (see Figures in Appendix \ref{appendix:best_fit_models}). In the second step, we excluded the CO region to avoid potential CO contamination from the circumstellar material \citep[as seen, for example, in ][]{Doppmann2005, Flores2020}. We anchored the projected rotational velocity to the best value found in the previous step and re-calculated the temperature, gravity, magnetic field strength, \textit{K} band veiling, and micro-turbulence. In both steps, we fitted the spectra with a Markov Chain Monte Carlo (MCMC) method using the \texttt{emcee} package \citep{Foreman-Mackey2013}. We ran a total of 50 individual chain ``walkers" advancing a total of 500 steps. Examples of how sensitive the stellar spectra are to parameters such as surface gravity and magnetic field strengths are depicted in Appendix \ref{appendix:gravity_measurements} and \ref{appendix:b_field_measurements}.

\section{Results} 
\label{sec:Results}

In Table \ref{table:Stellar_params}, we report all of the derived stellar parameters for the 40 stars. The measured \textit{K} band temperatures range from about 3000~K to 4400~K, with a median \textit{K} band temperature value of 3600~K, magnetic field strengths range from 0.7 to 3.2 kG with a median value of 1.8 kG, and surface gravity values span the range of 3.1 to 4.3 in log of $\rm cm \, s^{-2}$, with a median value of 3.8 in log of $\rm cm \, s^{-2}$. The parameter values are reported as the median  of the MCMC distributions while the uncertainty in the stellar parameters are reported as 3$\sigma$ values (see example in Appendix \ref{appendix:cornerplot}).

\subsection{Optical and IR temperature differences}
\label{subsec:starspots}

All the literature optical spectral types were transformed into temperatures using the temperature scales of \cite{Herczeg2014}. Our measured \textit{K} band temperatures are different, and almost always lower, than the optical temperatures. On the left panel of Figure \ref{fig:temperature_difference_optical}, we see a significant temperature difference between optical and IR measurements, which increases towards earlier optical spectral types. From the 37 stars with determined optical spectral types, only ROX~25 has an optical temperature measurement that is significantly lower than our \textit{K} band derived temperature ($\rm \Delta T = -285 \, K$). This intriguing case could be caused by a hidden multiplicity in the system, which might also explain the unusually weak magnetic field strength measured for this source ($B\sim0.7$~kG). Follow-up studies would be necessary to confirm this hypothesis. The right panel of Figure \ref{fig:temperature_difference_optical} shows a broad yet increasing stellar temperature difference as a function of the surface magnetic field strength. Put differently, stars with stronger magnetic fields have, on average, a larger temperature contrast between both measurements.

We assessed the statistical significance of the trends mentioned above using the Pearson correlation coefficient. We obtained Pearson coefficients of 0.60 and 0.59 for the temperature difference as a function of the optical temperature and as a function of the magnetic field strength, and P-values for testing non-correlations of $8.3 \times  10^{-5}$, and $1.3 \times  10^{-4}$, respectively. These tests shows that the variables are positively correlated and both relations are statistically significant. 

Additionally, we performed a linear fit to both relations. For the temperature difference as a function of optical temperature, we obtained $\Delta T = (0.36 \pm 0.08) \, T_{opt} - 1170 \pm 297~K$. This means that for stars hotter than 3200~K in the optical (M4 spectral type), there is, on average, a 36~K increase in $\Delta T$ for every 100~K increase in optical temperature.  For the temperature difference vs. magnetic field strength in kilogauss, we found $\Delta T = (206 \pm 44) \, B - 135 \pm 80$~K. We note that there is a significant scatter in both correlations. Thus, the linear relationships should be considered as an ensemble correlation rather than an object-by-object relationship.

Furthermore, these relationships are only valid for the temperatures considered in our sample and are expected to break at hotter and cooler temperatures. In Section \ref{sec:discussion}, we discuss physical mechanisms that could produce these effects in young stars and further speculate on the validity of these first-order fits outside the temperature range investigated here.

\subsection{Stellar masses from evolutionary models}

\begin{figure*}[ht!]
\epsscale{1.1}
\plotone{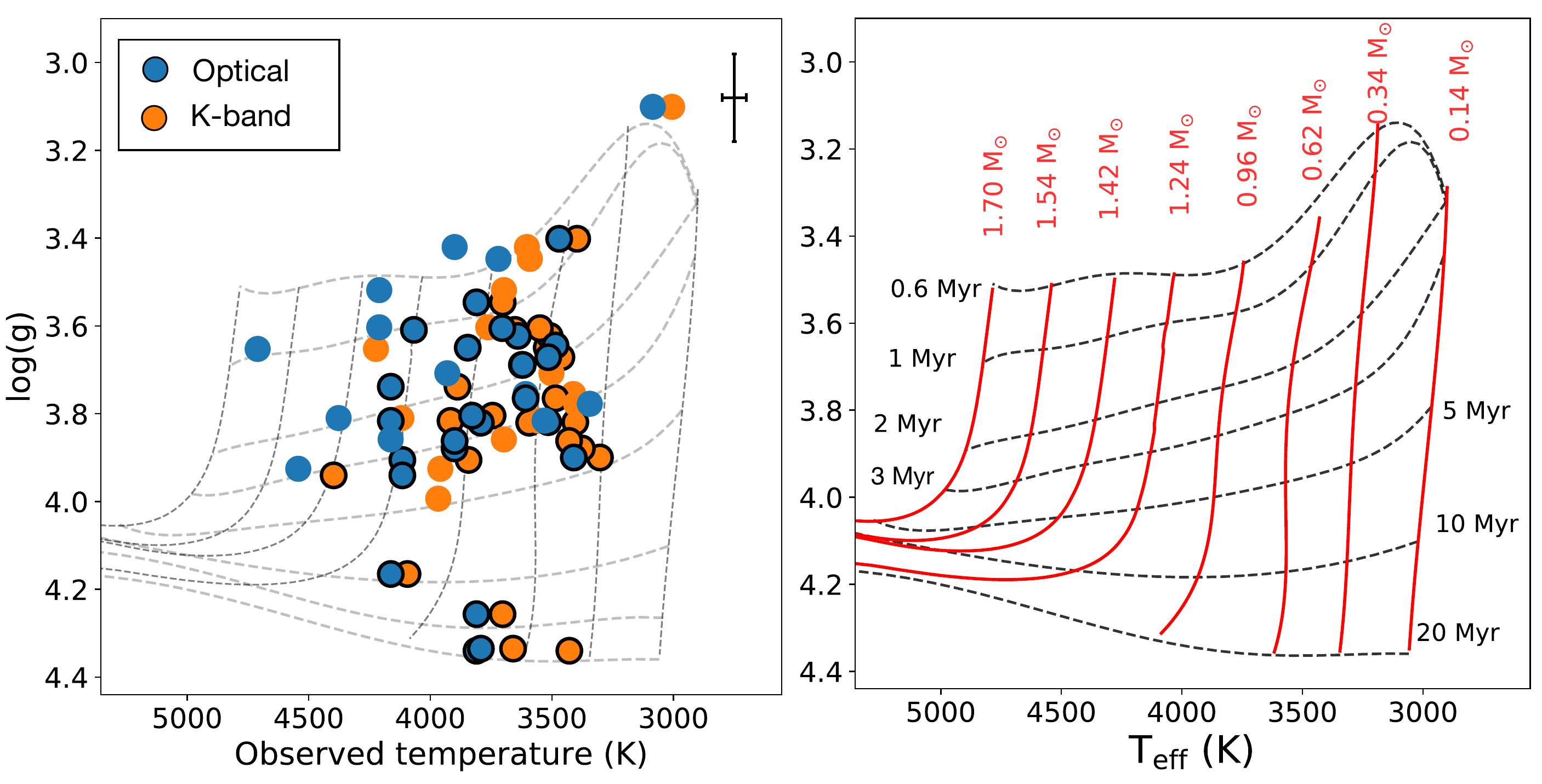}
\caption{Left panel: temperatures and gravities plotted in the HR diagram. Blue circles represent optical temperature measurement, while the orange ones correspond to \textit{K} band derived temperatures. Thick black contours demarcate stars with measured dynamical masses. Representative 3$\sigma$ uncertainties for the \textit{K} band temperature and gravity of  50~K and of 0.1~dex are shown on the upper right. Right panel: Feiden's magnetic mass tracks (red) and isochrones (black) are shown in a $\rm T_{eff}$ vs. gravity space. The scale on both panels is the same. \label{fig:Teff_compared_to_observed_temp}}
\end{figure*}

The effective temperature of a star is a mathematical construction that satisfies $T_{eff} = (L_\star/4\pi \sigma R_\star^2)^{1/4}$. The computation of effective temperatures in this form is difficult because either radii measurements or stellar luminosities over broad spectral ranges can be hard to obtain. Thus, it is customary to instead measure the temperature of a star from stellar spectra using a combination of atmospheric models and radiative transfer codes (often called spectroscopic temperature $T_{spec}$). The spectroscopic temperature is assumed to be a good representation of the effective temperature of a star regardless of the observed wavelength. 
Although there are technical differences for temperatures measured at different wavelengths (such as the knowledge of the atomic and molecular line lists, the characterizations of the instrument spectral profiles and responses, among others) proper calibrations of the stellar atmospheric models should provide consistent values. 

\cite{Flores2019} demonstrated that iSHELL $K$~band spectroscopic observations combined with the same radiative transfer code used here, provide accurate effective temperatures for main and post-main sequence stars (see Figure 6 of their paper). We thus argue that the observed temperature differences displayed in Figure \ref{fig:temperature_difference_optical} cannot be solely accounted for by uncertainties in the techniques used at the two different wavelengths, instead we propose that they reveal a true physical peculiarity of the stellar surfaces in low-mass young stars.

Furthermore, since young stars evolve almost isothermally during their first few million years of evolution \citep{Hayashi1961}, the mass of a star can be almost directly inferred from its effective temperature using `theoretical' HR diagram ($\log{g}$ vs. $T_{eff}$, see Figure \ref{fig:Teff_compared_to_observed_temp}). The problem of having two different temperatures for the same star, then translate into having also two discrepant stellar masses. 

To measure how different these two masses could be, we computed stellar masses using our gravities, the literature optical temperatures, the infrared temperatures, and the magnetic stellar evolutionary models from \cite{Feiden2016}. We chose these magnetic models because they provide more accurate stellar masses than other evolutionary tracks for young sources in the mass range of $ \rm M_\star \lesssim 1.3 \, M_\odot$ \citep{Simon2019,Braun2021}. We use a `theoretical' HR diagram ($\rm log{(g)}$ vs. $\rm T_{eff}$) instead of the `observational' HR diagram ($\rm L_{\star}$ vs. $\rm T_{eff}$) because a) gravities are directly probed in the stellar spectra and b) because it circumvents multiple sources of biases introduced by the measured luminosity such as extinction, variability, and multiplicity.

\begin{figure*}[ht!]
\epsscale{1.0}
\plotone{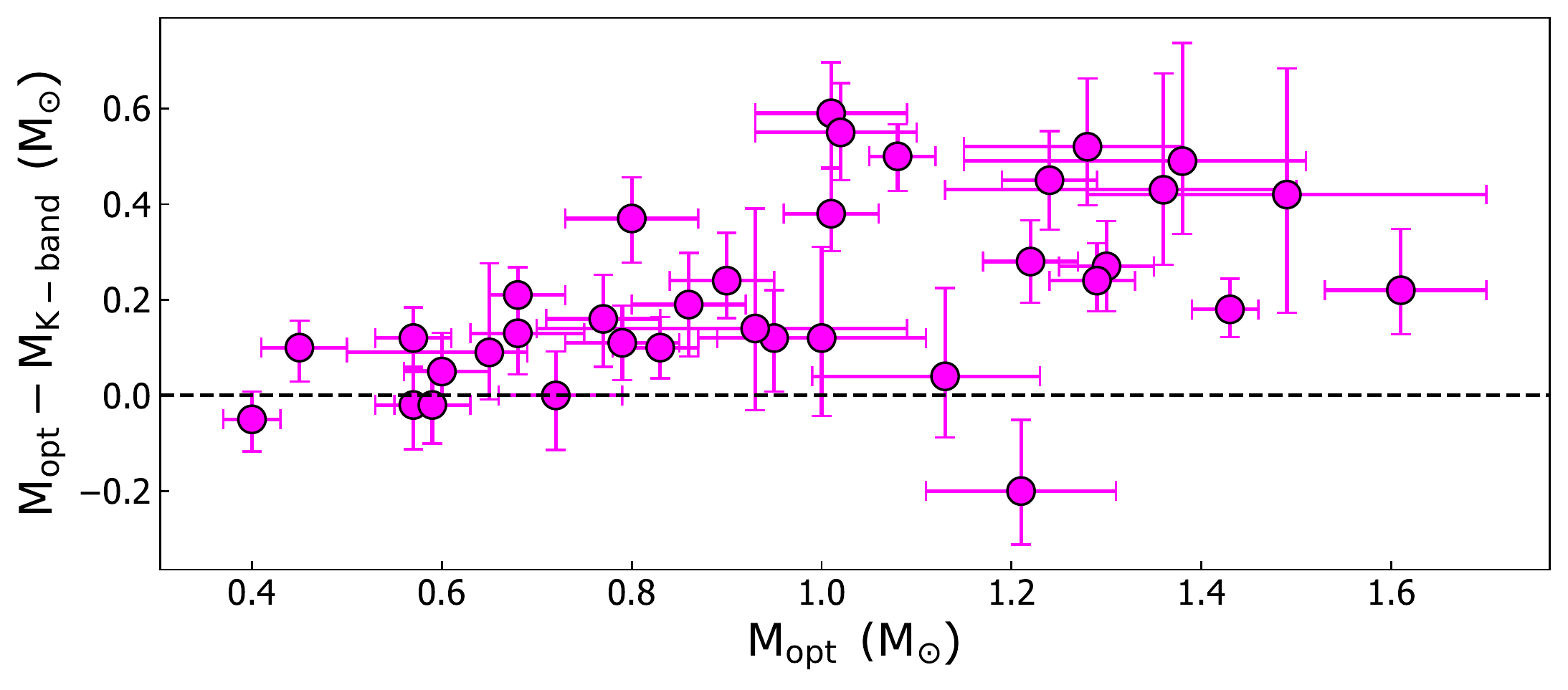}
\caption{Difference between masses calculated from \cite{Feiden2016} magnetic models using literature optical and our K band infrared temperatures. The gravities were assumed to be the same for both measurements. Masses were only calculated for the 35 sources that completely lie within the grid of stellar evolutionary models. \label{fig:Optical-mass-vs-IR-mass}}
\end{figure*}

In  Figure \ref{fig:Teff_compared_to_observed_temp}, we plot our sample of stars with optical literature measurements and infrared temperatures using our derived gravities in both cases. Since optical temperatures are generally hotter than the IR-derived ones, stars with optical temperatures lie on higher mass tracks than stars with \textit{K} band temperatures. From the figure, we see that only 35 of the stars in our sample completely lie within the grid of stellar evolutionary models, and of those stars, on average, masses obtained from optical temperatures are $28 \pm 2 \, \%$ higher than the stellar masses calculated from infrared temperatures.

Trends in the difference between stellar masses derived from optical temperatures vs. infrared temperatures are depicted in Figure \ref{fig:Optical-mass-vs-IR-mass}. Not surprisingly, it shows that stellar masses derived from optical temperatures are almost always higher than the ones obtained from infrared measurements and this difference increases for stars with higher optically-derived masses.


\subsection{Dynamical vs. spectroscopic masses}
\label{sec:tempr_masses_vs_dynamical}

The natural question is then, does the optical or infrared temperature better correspond with the mass of the star? To address this, we focus on a sub-sample of our young stars that have dynamical masses derived from CO and CN observations obtained with the Atacama Large Millimeter/submillimeter Array (ALMA) and the Plateau de Bure Interferometer (PdBI) \citep{Schaefer2009,Guilloteau2014,Simon2019}. In the following analysis, we only consider sources that have dynamical mass measurements with a precision better than $20 \%$. After this selection criterion, 24 sources were left within the mass range of 0.3~$\rm M_\odot$  to 1.3~$\rm M_\odot$. 

\begin{figure*}[ht!]
\epsscale{1.0}
\plotone{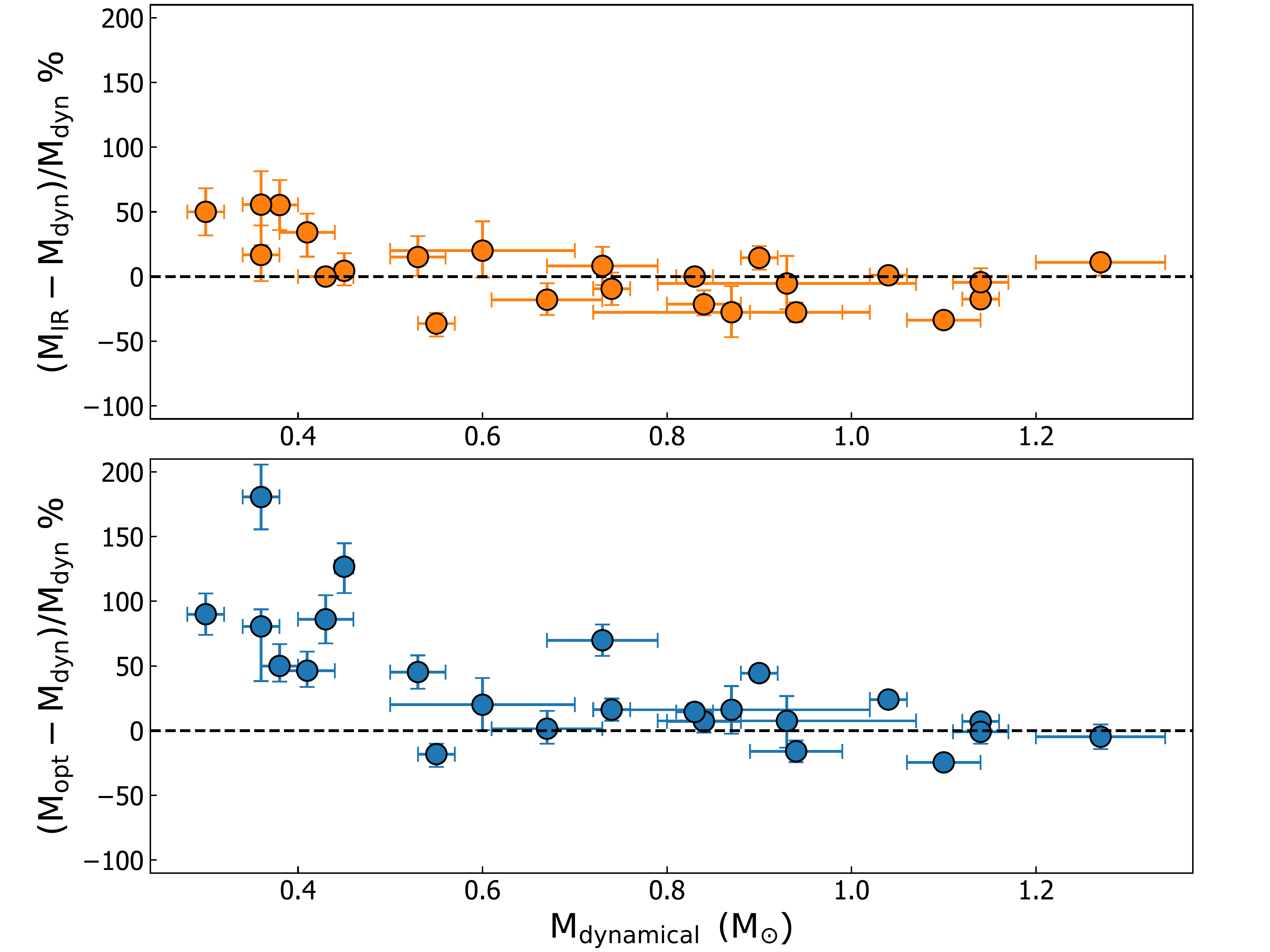}
\caption{Both panels show the difference between model-derived stellar masses and dynamical masses as a percentage of the dynamical masses for 24 young stars. The upper panel shows the result for masses derived using our \textit{K} band temperatures. The bottom panel displays the results for masses calculated using optical literature temperatures. The \textit{K} band derived masses are not only more precise but also more accurate than the optical masses. \label{fig:masses}}
\end{figure*}

Figure \ref{fig:masses} shows the difference between the spectroscopic derived masses and the dynamical masses calculated from interferometric observations as a percentage of the dynamical masses. \textit{K} band and optically derived masses reproduce reasonably well the dynamical masses of sources with $\rm M_{dyn} \gtrsim 0.5 \, M_\odot$. In fact, in this mass range, the mean percentage difference for the IR-derived masses  is $-8\pm 4 \,\%$ with a standard deviation of $18 \, \%$. The optical measurements in the same range, produce a mean percentage difference of $12\pm 6 \,\%$, with a standard deviation of $24 \, \%$. For lower stellar masses $\rm M_{dyn}<0.5 \, M_\odot$, both spectroscopic temperatures produce masses in excess of the measured dynamical masses, with the optically-derived masses performing significantly worse than the infrared ones. The infrared derived masses, over-predict the stellar masses by $31\pm 8 \,\%$, while the optical masses over-predict the stellar masses by $94\pm 16 \,\%$. 

Considering the full mass range ($\rm 0.3 - 1.3 \, M_\odot$), the infrared-derived masses have a mean percentage difference of $4\pm 5 \,\%$ with a standard deviation of $26 \,\%$. The optical masses, in turn, have a mean percentage difference of $36\pm 8 \,\%$ with a standard deviation of $48 \,\%$. Although neither model-derived mass is a perfect representation of the dynamical mass of the stars, this analysis implies that masses derived from \textit{K} band observations are not only more precise than masses derived from optical, but they are also more accurate.

\section{Discussion} \label{sec:discussion}

\subsection{Starspots on young stars}
\label{subsec:origin_temp_differences}
Our interpretation of the temperature differences between optical and infrared observations is that they are caused by thermal inhomogeneities on the surface of the young stars, i.e., cold and hot spots. The presence of spots is supported by the fact that all the observed young stars have strong ($\sim$kG) magnetic fields, and the strength of the magnetic fields positively correlates with the observed temperature difference. Although this alone cannot prove a causation between magnetic fields and starspots, we note that temperature inhomogeneities in main and post-main sequence stars are intrinsically linked to magnetic effects. This is the case, for example, for the Sun, the well studied RS CVn Stars \citep[see reviews of][]{Berdyugina2005,Strassmeier2009}, the BY Drac type stars \citep{Bopp1973,Anderson1976}, and the FK Comae stars \citep{Bopp1981,Jetsu1993}.

\begin{figure}[ht!]
\epsscale{1.1}
\plotone{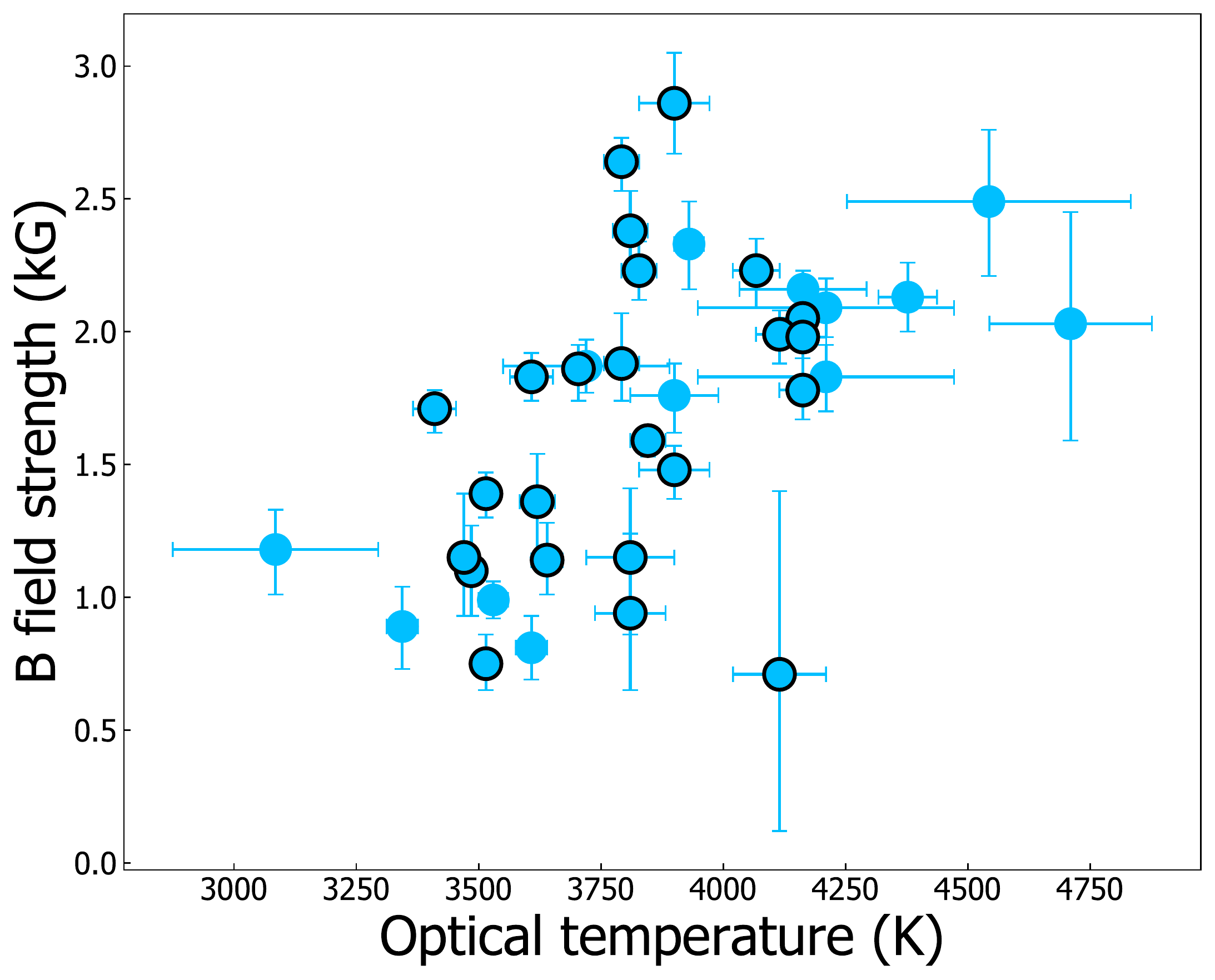}
\caption{Magnetic field strengths as a function of literature optical temperatures. The increasing magnetic field strengths as a function of optical temperature are opposite to a trend found by \cite{Sokal2020}. \label{fig:B-fields_vs_optical_temp}}
\end{figure}

As shown in Figure \ref{fig:B-fields_vs_optical_temp}, our measured magnetic field strengths positively correlate with literature optical temperatures. This finding, however, is in contradiction with results found in the previous study of \cite{Sokal2020} who compiled magnetic field strengths of classical T Tauri stars (cTTS) and weak-line T Tauri stars from different authors \citep[e.g.,][]{Johns-Krull2007,Yang2011, Lavail2017}, and plotted them against their optical temperatures. With a sample of 28 cTTS, 12 wTTS, and one Class~I source, they found an anti-correlation between the measured magnetic field strengths and the optically measured temperatures, which was attributed to a possible evolutionary change in the stars. 

To check for biases in the measured magnetic field strengths, we compared the nine stars in common between the compilation of \cite{Sokal2020} and our survey. These nine sources have magnetic field strengths ranging from 0.7 to 2.4~kG, and differ by $\sim13\%$ (median difference) with our B-field strengths, with the higher values presented in \cite{Sokal2020}.

We propose two explanations for the observed differences; a) a real variations in the B-field strength of the stars over decade-long periods. The abovementioned sources were observed by \cite{Johns-Krull2007} between 1996 and 2004, leading to a 13-24 years difference with our observations (see Section \ref{sec:observations}). Although no study has proven magnetic variability over such a long timespan using a consistent technique, \cite{Lavail2019} measured magnetic field strength variability of a sample of seven young stars over a period of 2-5 months. They found a typical variation of the order of 0.3~kG (or $\sim$ 10\%), which could suggest that magnetic field variability is possible over longer timespans. b) another possible explanation for the discrepant magnetic field strengths is the difference of the instruments, atmospheric lines, radiative transfer codes, and underlying assumptions used in each study. The abovementioned nine sources were observed with CSHELL which allowed to observed three wavelength regions in the $K$~band: two Ti-line regions, and one region that targets part of the CO overtone \citep{Johns-Krull2007}. The wavelength range of the CSHELL observation was more than ten times smaller than what we acquired with iSHELL and also at half the spectral resolution. Furthermore, stellar parameters such as temperature, gravity, and rotational velocity were not self-consistently derived but adopted from previous studies.

Since the two modeling techniques and data quality are significantly different (due to the rapid technological change in detectors and computers), we lean towards the latter explanation rather than the former. However, we do not entirely discard real magnetic field strength variations as a possibility. In any case, given that our results are opposite of the trend found in \cite{Sokal2020} and because both studies contain a similar number of sources, only a follow-up survey with a larger number of stars can confirm or support either trend.

\subsection{Temperature difference and stellar structure}
\label{subsec:stellar_structure}

The positive correlation between the optical temperature and the temperature difference (shown in Figure \ref{fig:temperature_difference_optical}) is not expected to extend indefinitely in either temperature direction. Towards the cool end, at about an optical temperature of $\sim3200$~K, stars should manifest a negligible optical to infrared temperature difference. One way to produce this is that the temperature of the cool spots on the stellar surface approaches the temperature of the warm photosphere. For example, they might follow a fixed $ T_{\rm spot}/T_{\rm phot}$ ratio, similar to what is seen in RS CVn stars  \citep[e.g.,][]{Catalano2002,Frasca2008}, which would reduce the overall contrast between optical and IR observations. This alone, however, would not completely explain the observed pattern towards the cool end, as it would also be necessary that hot spots become less important. 

For stars on the hot end, the temperature differences could keep increasing as a function of their effective temperature. These differences, however, should eventually decline as the convective envelope of the stars become shallower and the surface magnetic fields of the stars weaken. \cite{Lavail2017} measured the magnetic field strengths of a sample of six intermediate mass T Tauri stars (IMTTS) and found evidence that their values were, on average, lower than what has been measured for low-mass T Tauri stars.

The internal structure and thus magnetic field of the stars is expected to strongly depend on the stellar temperature and age \citep[see e.g.,][]{Gregory2012,Villebrun2019}. 
This means that observing earlier type young stars ($T_{opt}>5000$~K) in both the infrared and optical could provide direct observational evidence of convective layers in stars becoming shallower, thus allowing us to understand the evolution of stellar internal structure better and posing constraints on evolutionary models.

\subsection{Evolutionary models and dynamical masses}

In Section \ref{sec:tempr_masses_vs_dynamical}, we showed that the infrared-derived masses are more accurate and precise than the ones derived from optical temperatures. However, the underlying assumption in our analysis is that the magnetic models reported by \cite{Feiden2016} are correct. It could be argued that this analysis shows that the magnetic models from Feiden provide incorrect masses for a given temperature. Although we cannot prove otherwise, we can rely on previous studies that have shown the reliability of \cite{Feiden2016} magnetic models. \cite{Simon2019} compared magnetic vs. non-magnetic models from different authors \citep{Baraffe2015, Feiden2016} and determined that magnetic models produce better results for young stars in a similar mass range to what is investigated here. Recently, \cite{Braun2021} also tested various stellar evolutionary models against ALMA dynamical masses of young stars and concluded that the \cite{Feiden2016} magnetic models provide the best estimate of the stellar masses in the mass range of 0.6 to 1.3 $\rm M_{\odot}$. For sources less massive than 0.6 $\rm M_{\odot}$, \cite{Braun2021} found that the \cite{Feiden2016} models overestimate the stellar masses. Our study confirms these findings but also suggests that this problem is more severe when using optical temperatures \citep[as was done in][]{Braun2021}, than when using infrared-derived temperatures of the young stars.

There are two other aspects that must be kept in mind regarding our analysis. First, we assumed that our measured gravities can be used in combination with literature optical data to derive stellar masses. This assumption makes sense from a physical perspective, as starspots should be in hydrostatic equilibrium with the ambient photosphere, but in practice, the gravities and optical temperatures were calculated independently from different datasets, instruments, and techniques. Second, the dynamical mass values adopted in our computations could be more uncertain than what was originally reported by \cite{Simon2019}. 
\cite{Braun2021} used archival ALMA data to measure dynamical masses of young stars in Taurus. Their sample overlapped for 8 sources with the Taurus sources analyzed by \cite{Simon2019}, and discrepancies of dynamical mass measurements were as high as $46\%$. As explained in \cite{Braun2021}, these stellar-mass differences mainly arise due to the difference in the adopted disk inclination angles (since stellar masses scale as $M_\star \propto 1/ \sin^{2}{i}$). 

Future gas observations of these disks at higher resolution and $S/N$ could help to decide which measurements we should favor. 

\section{summary and conclusion} \label{sec:conclusion}

We obtained iSHELL \textit{K} band observations of 40 pre-main-sequence sources from Taurus-Auriga and Ophiuchus star-forming regions. We used a magnetic radiative transfer code to derive parameters of the stars such as temperature, gravity, and magnetic field strength. We compared our derived values against optical literature temperatures and used stellar evolutionary models to test whether optical or infrared temperatures should be used to derive stellar masses of young stars. A summary of our findings are as follows:

\begin{itemize}
    \item Our \textit{K} band derived temperatures are almost always lower than optical temperatures from the literature, and this difference increases as a function of the optical temperatures. Additionally, the observed difference between optical and infrared temperatures increases as a function of the magnetic field strengths of the stars.
    \item We interpret that the temperature differences between optical and infrared observations are caused by hot and cold spots on the stellar surface. This interpretation is supported by the fact that, on average, stars with stronger magnetic fields have larger temperature differences and is also reinforced by the fact that magnetic activity and thermal inhomogeneities are intrinsically linked in a variety of main and post-main sequence stars.
    \item Using the \cite{Feiden2016} magnetic models, we computed the stellar masses for 35 young sources using both optical and infrared temperatures. The mean spectroscopic mass calculated from optical temperatures was $28\pm2\%$ higher than from the \textit{K} band temperatures. This implies that the choice of optical vs. infrared temperatures when using stellar evolutionary models is important.
    \item We used a sub-sample of 24 stars with masses calculated from ALMA and PdBI to explore whether optical or infrared temperatures better reproduce the dynamical stellar masses. In the mass range of 0.3~$\rm M_{\odot}$ to 1.3~$\rm M_{\odot}$, we found that the IR temperatures produce more precise and accurate stellar masses than optical temperatures.
\end{itemize}

In conclusion, we have shown that magnetically-induced starspots can significantly alter the temperature measured on young stars. Although it is common to use optical temperatures to infer stellar masses, we have demonstrated that infrared temperatures should be preferred in the mass range of 0.3 to 1.3 $M_\odot$. Future studies on a broader range of stellar masses will be necessary to test whether our findings also apply to higher mass young stars.



\acknowledgments
We thank the referee for an insightful review which helped to improve this paper. CF and MSC acknowledge support from the NASA Infrared Telescope Facility, which is operated by the University of Hawaii under contract 80HQTR19D0030 with the National Aeronautics and Space Administration. GD acknowledges support from NASA grants NNX15AC89G and NNX15AD95G/NExSS as well as 80NSSC18K0442. The technical support and advanced computing resources from University of Hawaii Information Technology Services – Cyberinfrastructure, funded in part by the National Science Foundation MRI award $\# 1920304$, are gratefully acknowledged. We are particularly grateful to the
iSHELL team for all the help provided during the observation and data reduction process. This research has made use of NASA's Astrophysics Data System Bibliographic Services. This research has made use of the SIMBAD database, operated at CDS, Strasbourg, France. This work has made use of the VALD database, operated at Uppsala University, the Institute of Astronomy RAS in Moscow, and the University of Vienna.

\vspace{10pt}
\newpage

\input{stellar_param-Feb20-2021.tex}
\input{Stellar_masses.tex}
%

\vspace{5mm}
\facilities{IRTF, ALMA, PdBI}


\software{astropy \citep{2013A&A...558A..33A}
          MoogStokes \citep{Deen2013}, 
          emcee \citep{Foreman-Mackey2013}, 
          Spextools \citep{Cushing2004}, 
          xtellcor \citep{vacca2003}
          }


\newpage

\appendix

\section{Individual Stellar Spectra and Best Fit Models} 
\label{appendix:best_fit_models}
\begin{figure*}[ht!]
\epsscale{1.0}
\plotone{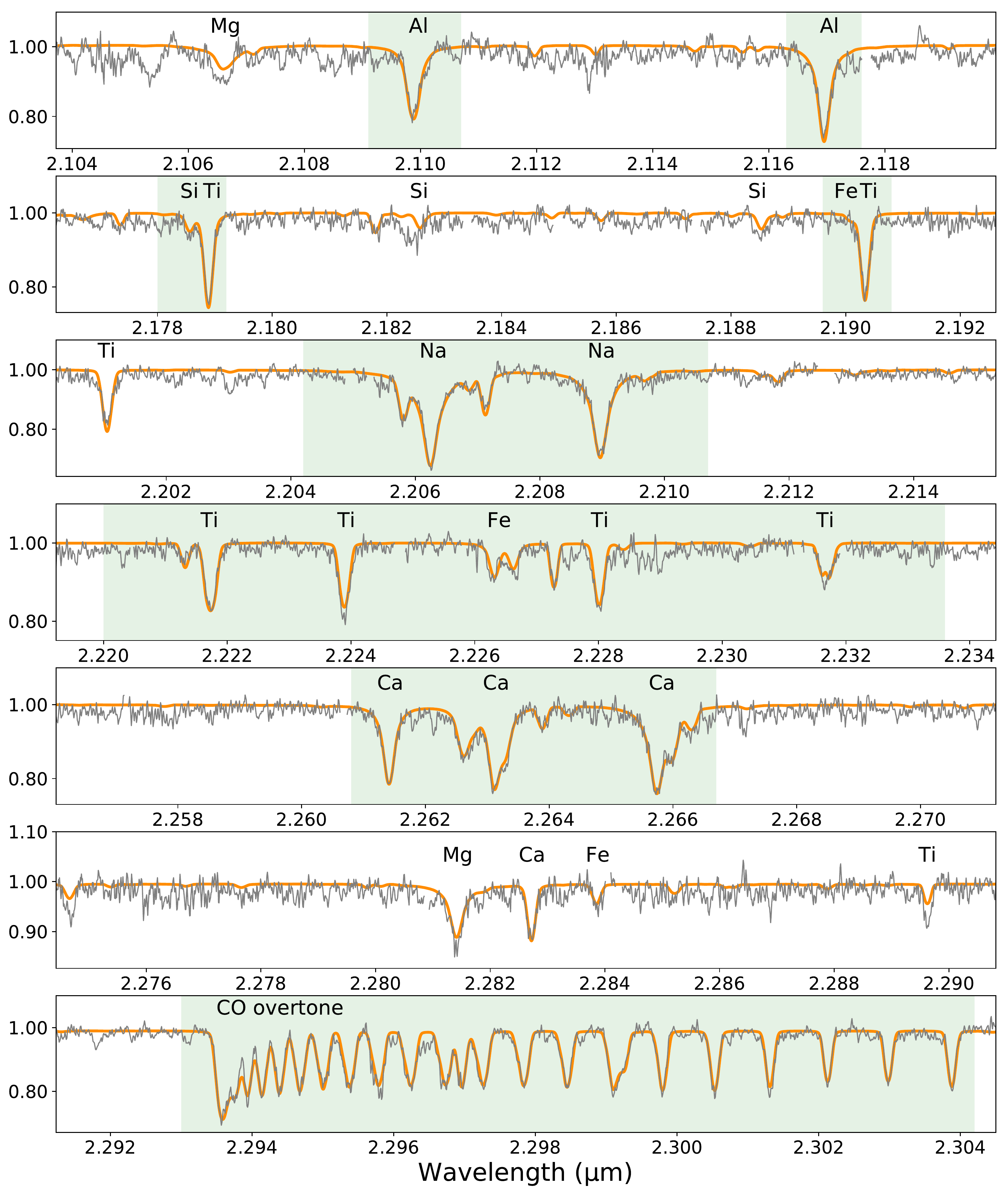}
\caption{Comparison between the spectrum of CY~Tau (in gray) and its best fit model (in orange). The seven panels show regions with the strongest photospheric absorption in lines. Green faded boxes denote the regions used in the determination of stellar parameters. In each panel, some of the absorption lines are labeled.  \label{fig:CY-Tau}}
\end{figure*}

\begin{figure*}[ht!]
\epsscale{1.0}
\plotone{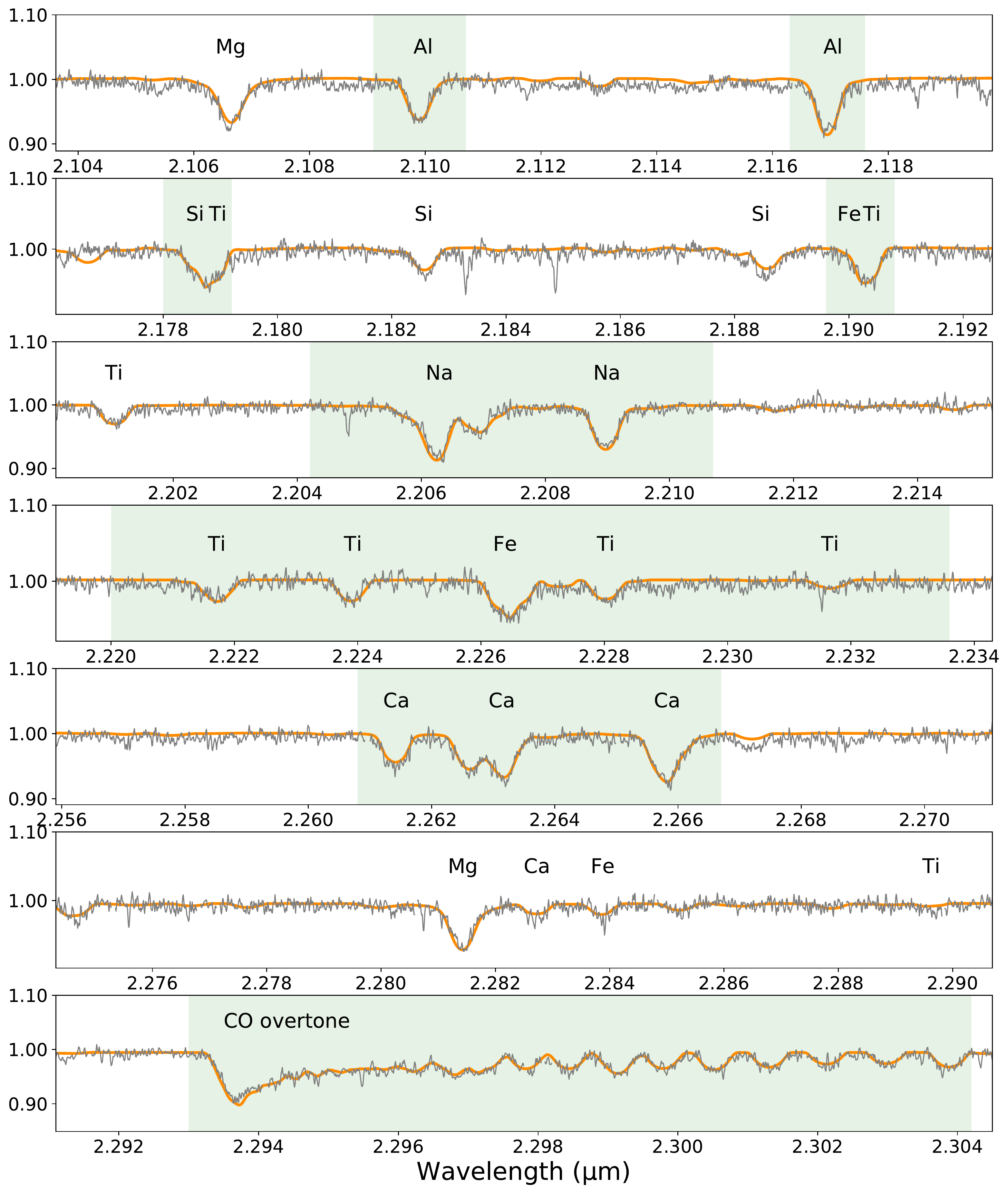}
\caption{Same as before but for DoAr 43SW \label{fig:DoAr43SW}}
\end{figure*}

\begin{figure*}[ht!]
\epsscale{1.0}
\plotone{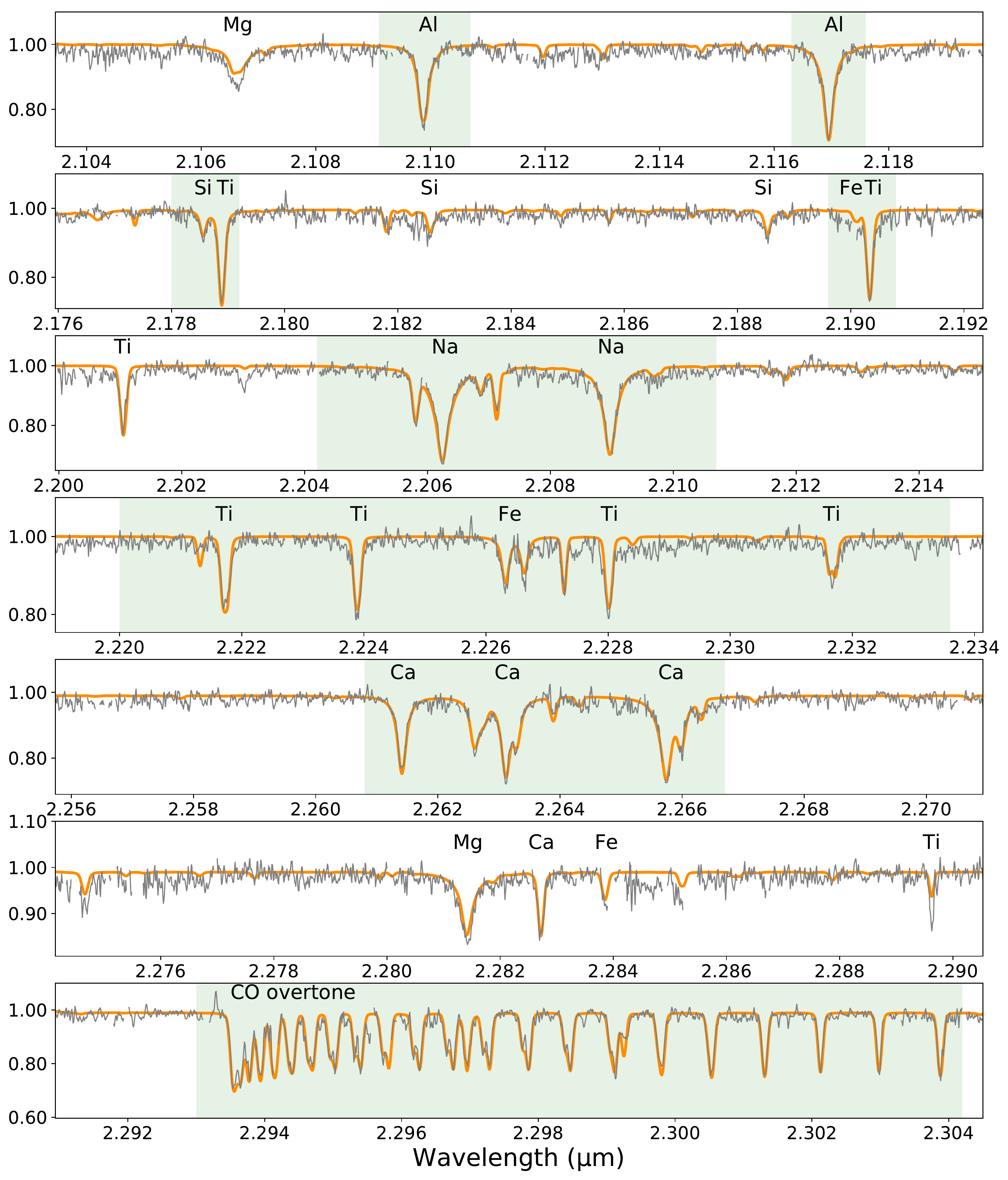}
\caption{Same as before but for FX Tau A. \label{fig:FX-Tau} The complete figure set of 40 images is available in the online journal.}
\end{figure*}

We present portions of the $K$~band spectrum of 3 sources and their best fit models in Figures \ref{fig:CY-Tau}, \ref{fig:DoAr43SW}, and \ref{fig:FX-Tau}. The figures for the remaining 37 stars can be accessed in the online journal as Figure sets. The plots display seven panels, each covering wavelength regions of about 150~\AA, that contain the strongest absorption lines in the spectra of the stars. Green faded boxes delimit the regions that were used to derive the stellar parameters of the stars (see Section \ref{sec:stellar_params}). For visual clarity the spectrum of the stars was smoothed with a 5-pixel wide Savitzky-Golay function of order 2. This smoothing was performed only for figure display and not during the data fitting process.

The three sources selected for the printed journal, were chosen to display a range in $K$~band temperature and projected rotational velocities. CY~Tau (shown in Figure \ref{fig:CY-Tau}) has a $K$~band temperature of $T_{K-band} = 3403^{+43}_{-21}$~K, and displays a ``forest" of molecular water absorption lines (not included in our models) which affect mostly, but are not limited to, the wavelength regions shortward of 2.2 $\micron$. At these cool temperatures (compare to the rest of the sample), the spectra of the stars are also characterized by a large line-depth difference between the Si and the Al lines at 2.179 $\micron$ (second panel from top to bottom), and weaker Fe lines at 2.2265 $\micron$ compared to the Ti lines at 2.222, 2.224, 2.228, and 2.232 $\micron$ (fourth panel). Hotter stars such as DoAr~43SW (shown in Figure \ref{fig:DoAr43SW}) with $T_{K-band} = 4223^{+64}_{-57}$~K, display no molecular absorption bands from the water molecule, a weaker ratio between Ti and Si at 2.179 $\micron$, and comparatively stronger Fe to Ti lines in the region of 2.221 to 2.232 $\micron$.

The differences in the projected rotational velocity of the sources can be assessed from almost any absorption line in the spectrum of the stars, but the unequivocal determination come from the width of the CO lines starting at  $\sim 2.3 \, \micron$, and the slope of the CO bandhead at 2.2937 $\micron$ (last panel). The CO absorption lines in the $K$~band are excellent indicators of the projected rotational velocity of a star as their width are not affected by changes in the magnetic field strength of the star, and very little affected by thermal and pressure broadening \citep[e.g.,][]{Johns-Krull2001, Doppmann2005}. In Figure \ref{fig:FX-Tau}, we show the spectrum of FX~Tau~A, which display well resolved CO spectral lines, an almost vertical CO overtone at 2.2937 $\micron$, and has a projected rotational velocity of $v\sin{i}=5.6^{+0.34}_{-0.19}$~$\rm km\,s^{-1}$. In comparison, the spectrum of DoAr 43SW (Figure \ref{fig:DoAr43SW}) shows broad absorption lines throughout the full $K$~band spectrum, a first CO overtone with a slope that deviates from a vertical line, and clearly rotationally broaden CO lines beyond 2.296 $\micron$. Concordantly, the projected rotational velocity measured for this star is of $v\sin{i}=34.6^{+1.13}_{-2.06}$~$\rm km\,s^{-1}$, which is largest velocity value measured in our sample.

\figsetstart
\figsetnum{1}
\figsettitle{Young Stellar Spectra}

\figsetend

\section{Gravity measurement}
\label{appendix:gravity_measurements}
Unlike other stellar parameters such as temperature, projected rotational velocity, and magnetic field strength, the change in surface gravity has a more subtle effect on the spectrum of the stars. However, given the high spectral resolution of our observations (R$\sim$50000), and the high quality of our data (S/N$>50$), we can make accurate and precise measurements of this stellar parameter.

\begin{figure*}
\gridline{\fig{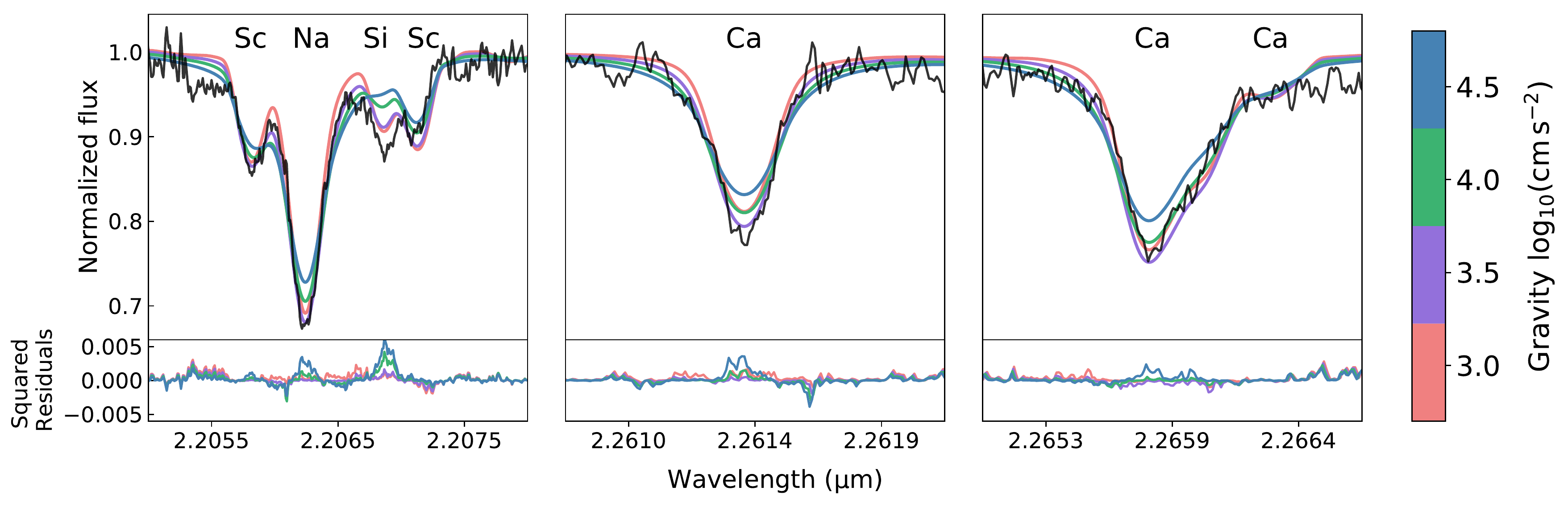}
{1.0\textwidth}{(a) GSS37W - $log{g}=3.45$}
   }
\gridline{\fig{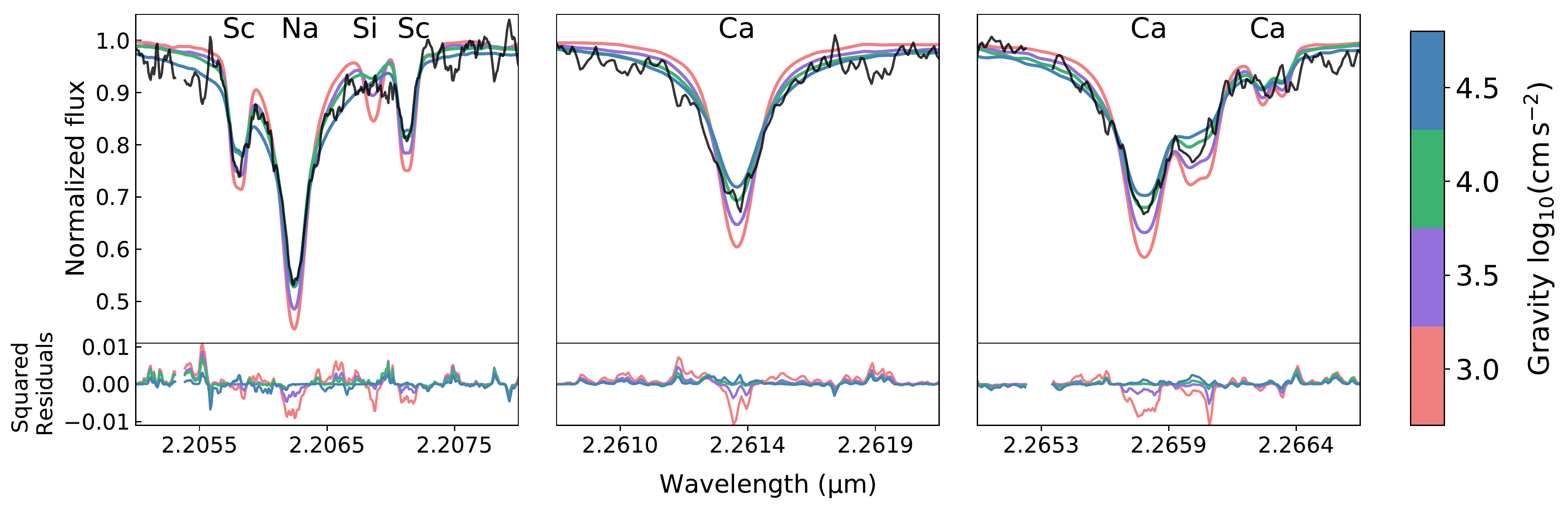}{1.0\textwidth}{(b) DM Tau - $log{g}=3.90$}
          }
\gridline{\fig{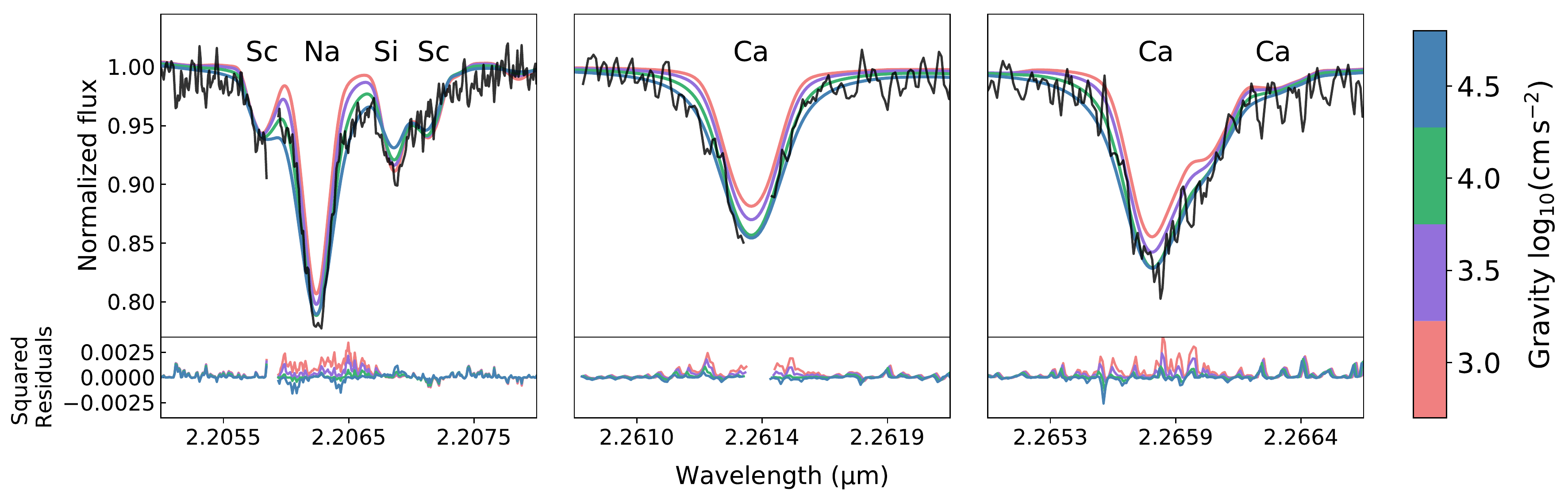}{1.0\textwidth}{(c) LkCa 15 - $log{g}=4.17$}}

\caption{Comparison between MoogStokes models with different gravities vs. the spectrum of (a) GSS37W, (b) DM Tau, and (c) LkCa 15, in the Na 2.206 $\micron$ (left), the Ca 2.261 $\micron$ (middle), and the Ca 2.265 $\micron$ regions (right). Residuals are squared as they better trace the chi-square statistic, but the signs are preserved for graphical purposes.\label{fig:logg_explanation}}
\end{figure*}

In Figure \ref{fig:logg_explanation}, we show that gravity changes of 0.5 dex in the spectrum of stars can be distinguished from the depth and strength of particular absorption lines. As example, we compared the spectrum of GSS~37W, DM~Tau, and LkCa~15 with gravities of $\log{g}= 3.45^{+0.13}_{-0.06}$, $3.90^{+0.02}_{-0.03}$, and $4.17^{+0.09}_{-0.17}$, respectively, against MoogStokes models with gravities ranging from 3.0 to 4.5 in $\log_{10}$ of $\rm cm \, s^{-2}$. For each source, all the other stellar parameters were fixed to their best fit value obtained in Section \ref{sec:Results}. We chose these three sources because they comprise a substantial range in gravities and $K$~band temperatures, and are relatively slow rotators $v\sin{i}<18$~$\rm km \, s^{-1}$. Three small spectral regions were chosen to exemplify the how changes in gravity result in better or worse fit to the data. A small subplot at the bottom of each wavelength region shows the squared residuals, but with the sign preserved, i.e, $sign(data-model) \times (data-model)^2$. The residuals were squared as they better approximate what chi-square calculates.

In the top panel, we see that the purple line ($\log{g}=3.5$) better matches the spectrum of GSS37~W (black line) in each wavelength region, and the second closest match is the pink line ($\log{g}=3.0$). These matches are consistent with the derived gravity for GSS37~W of $\log{g}=3.45$. The same situation happens in the middle panel, where the model that closest match the data for DM~Tau is the green line with a gravity of $\log{g}=4.0$. For the bottom panel, the green and blue lines (of $\log{g}=4.0$ and $\log{g}=4.5$, respectively) closely reproduce the data for LKCa~15, which has a measured $\log{g}=4.17$. In these panels, regions that allow to better discriminate between different gravities are the depth and wings of the Na lines at 2.206 $\micron$, and the strength of the Ca line at 2.614 $\micron$ and Ca doublet at 2.2660 $\micron$.

For additional information on the precision and accuracy our derived gravities, we refer the reader to Figure 5 of \cite{Flores2019}, where we compared our measurements against literature gravities of main and post-main sequence stars. It is worth emphasizing, that the modeling approach described in Section \ref{sec:stellar_params} does not only use the lines shown in Figure \ref{fig:logg_explanation}, but all the information of the green boxes displayed in Figures \ref{fig:CY-Tau} to \ref{fig:FX-Tau}.

\section{Magnetic field measurement}
\label{appendix:b_field_measurements}
\begin{figure*}[ht!]
\epsscale{1.1}
\plotone{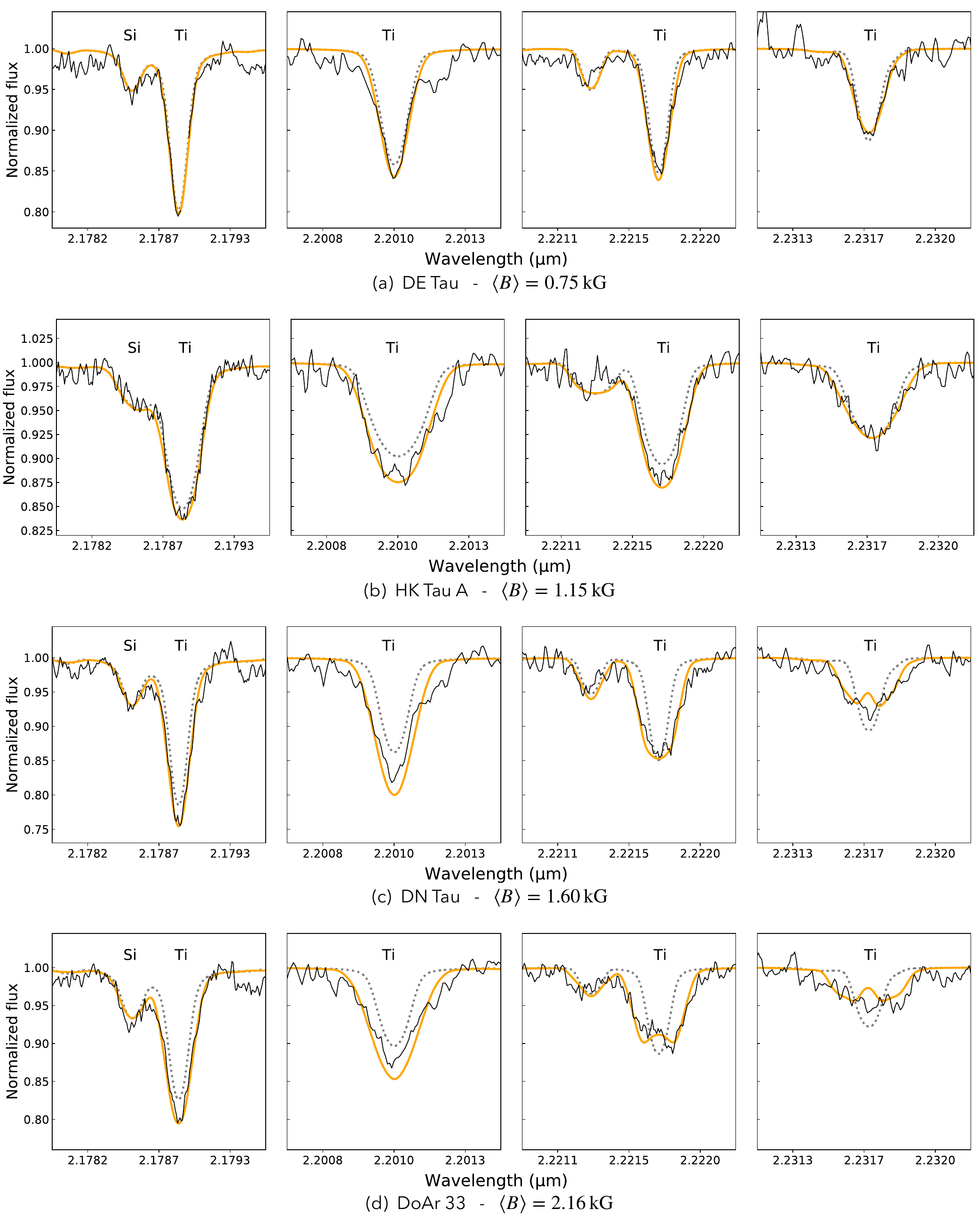}
\caption{Demonstration of the effects of magnetic fields on the spectra of DE~Tau, HK~Tau~A, DN~Tau, and DoAr~33. Four panels that include the highly magnetically sensitive Ti lines are shown in the figure. The spectra of the stars are displayed in solid black, the best fit synthetic models are shown in orange, and comparative synthetic models with null field strengths ($\langle B \rangle=0$ kG) are displayed as dotted black lines. In all cases, even for the weakest magnetic field strength detected, there are visible differences between the magnetic and null field models. \label{fig:B_field_explanation}}
\end{figure*}

One of the fundamental stellar parameters presented in this work is the magnetic field strengths of the stars. It is key for our interpretation of the temperature differences as caused by magnetically induced starspots. We therefore show in Figure \ref{fig:B_field_explanation} examples of how the magnetic field strength of the stars changes their spectra. We selected four stars with magnetic field strengths of $\langle B \rangle$ = 0.75, 1.15, 1.60, and  2.16 kG because they are a representative sample of the values found for the rest of the sources. The $K$~band temperature range of the selected sources varies between 3450~K and 3700~K, and the projected rotational velocities change between 10~$\rm km \, s^{-1}$ and 22~$\rm km \, s^{-1}$. Although the limited sample size makes it hard to select stars with similar stellar parameters, we emphasize that the plots shown in the appendix section are just meant to be a demonstration of our method. In the following paragraphs, we focused on atomic lines with strong magnetic sensitivity to illustrate that it is possible to detect magnetic fields as weak as 0.7~kG, and how line shapes can be significantly altered under the presence of a 2.0~kG field strength.

In Figure \ref{fig:B_field_explanation}, we display four sub-figures that correspond to small portions of the spectrum of DE Tau, HK Tau A, DN Tau, and DoAr 33. The sources were sorted from weakest to strongest magnetic field strengths, the stellar spectra is displayed in solid black, the best fit model is shown in orange, and a comparative spectrum with a null magnetic field strength is displayed as dotted and thinner black line. 

For DE~Tau (top row), we see almost no difference between the null magnetic field strength model (dotted line) and the best fit model with $\langle B \rangle = 0.75$~kG in the first panel, and only a small difference in the last three panels. For this source, the main observed difference is the width of the Ti lines for panels three and four, and a slight line depth difference in the second panel. The fact that we only see little differences between the models and the observations for this source, reflects that our method is approaching the detection limit. Indeed in \cite{Flores2019}, we used observations of non-magnetic giant stars to infer that the minimum magnetic field strength we can measure is $\langle B_{limit} \rangle = 0.3$~kG

The case of HK~Tau~A (second row) is similar to DE~Tau, where the null magnetic field model is not too different from the best fit model for this source. The largest difference for this star is seen for the Ti lines at 2.2010 and 2.2217 $\micron$ (second and third panel). For last two stars DN Tau and DoAr 33, which have the strongest magnetic fields, the models with zero magnetic fields and the best fit models are substantially different in all four panels. In particular, the Ti line at 2.2317 $\micron$ (fourth panel) shows a much broader profile, which is reproduced by the model through magnetic splitting of the line (Zeeman splitting). Although the S/N of the observations (or perhaps the simplistic one magnetic component assumption in our models) does not allow us to see the line splitting, the null-field model with same stellar parameters (temperature, gravity, rotational velocity, etc.) fails to reproduce both the width and depth of the Ti at 2.2317 $\micron$. In the case of DoAr 33, even the Ti line at 2.2217 $\micron$ splits (third panel), which indicates that the magnetic field of the star is significantly stronger than the other sources displayed here.


\section{Stellar parameter uncertainty and degeneracy}
\label{appendix:cornerplot}
\begin{figure*}[ht!]
\epsscale{1.}
\plotone{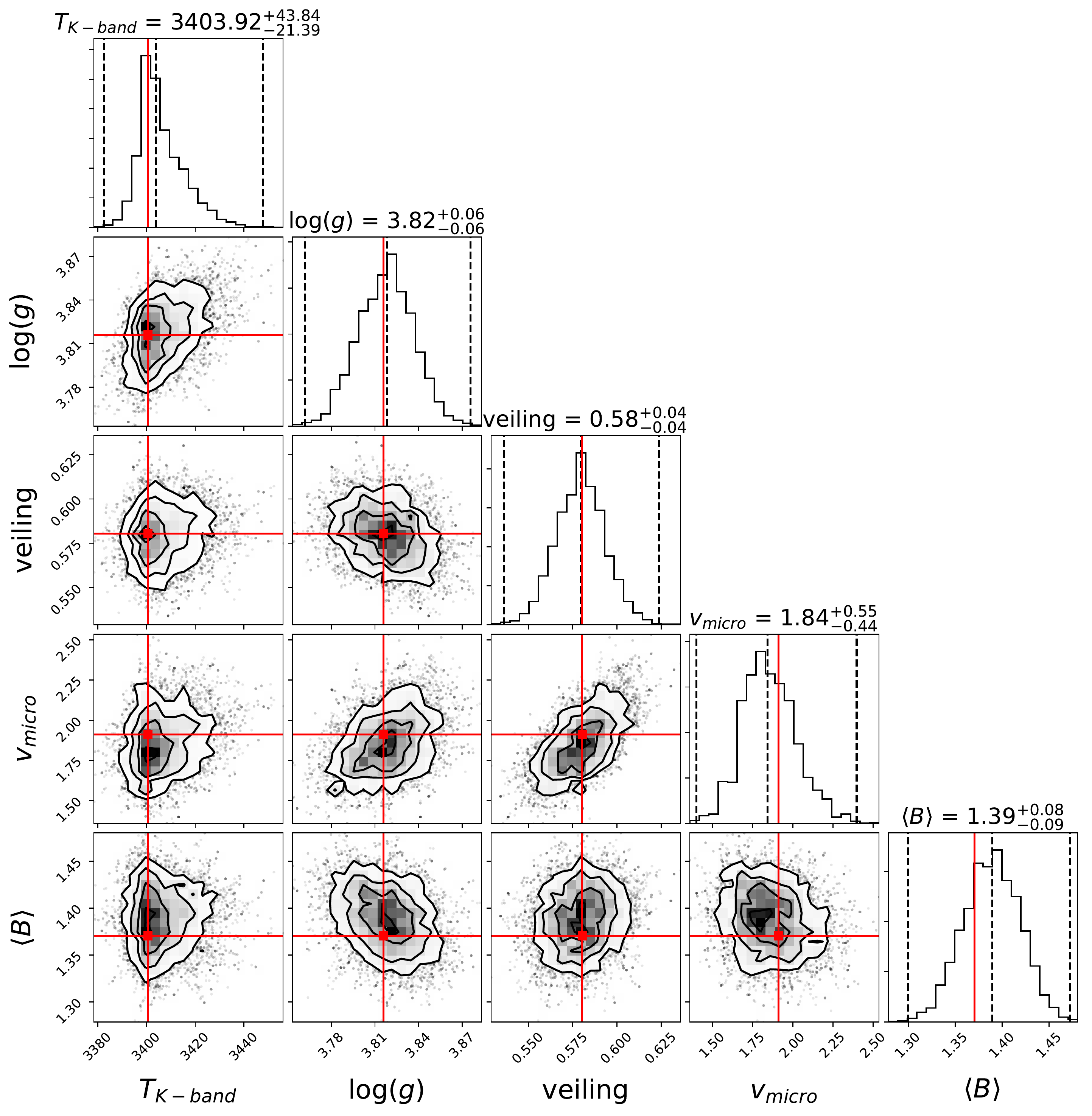}
\caption{We present the results for the second MCMC run for CY Tau (rotational velocity is not included). The corner plot shows only 50\% of the models, i.e., about 12500 models. The central black lines correspond to the median of the distributions, while the black lines on the side are the 3$\sigma$ uncertainties. The specific best fit model is traced by a red line. \label{fig:cornerplot}}
\end{figure*}

In Figure \ref{fig:cornerplot}, we present the corner plot for CY Tau made from its second MCMC run. The corner plot for the rest of the sources can be accessed in the online journal as Figure sets. As explained in Section \ref{sec:stellar_params}, the second MCMC step does no include the CO region (starting at $\sim2.293\, \micron$) neither it fits the $v\sin{i}$ parameter. For every source, the corner plots only include the last 50\% of the $\sim$50,000 models. We visually assessed the convergence of the MCMC runs, which typically happens within the first 20$\%$ of each run. There are small correlation between some stellar parameters, for example, the $v_{micro}$ and the veiling parameters are weakly correlated for this source. This makes sense, as both of these parameters are known to affect the depth of the lines. There are also some minor asymmetries in the marginalized two dimensional distribution for $v_{micro}$ vs. $\log{g}$, and for $\langle B \rangle$ vs.  $\log{g}$. For some of the stars the $T_{K-band}$ and the $\log{g}$ can also weakly correlate. The values quoted on top of the 1-D histograms correspond to the median of the distribution (central black dashed line), the side black dashed lines represent the 3$\sigma$ uncertainties. Red lines point to the specific best fit model of the MCMC chain. This best-fit model is well within the 3$\sigma$ uncertainty for each stellar parameter in every case.

\figsetstart
\figsetnum{2}
\figsettitle{Cornerplots of MCMC runs}

\figsetend

\newpage

\newpage
\bibliography{sample63}{}
\bibliographystyle{aasjournal}



\end{document}

%% file: stellar_param-Feb20-2021.tex
\begin{deluxetable*}{lcccccc}
\tabletypesize{\scriptsize}
\tablecaption{Stellar parameters of young stars derived from \textit{K} band. \label{table:Stellar_params}}
\tablehead{
\colhead{Source} & \colhead{$\rm T_{\rm K-band}$} & \colhead{log(g) } & \colhead{$\rm r_K$} &\colhead{$v_{micro}$}  & \colhead{$\langle \rm B \rangle $}   & \colhead{$v\sin(i)$}\\
\colhead{name} & \colhead{(K)} & \colhead{} & \colhead{} &  \colhead{(km s$^{-1}$)} & \colhead{(kG)}  & \colhead{(km s$^{-1}$)} }
\startdata
AA Tau    & $3593^{+61}_{-76}$ & $3.82^{+0.08}_{-0.11}$ & $0.99^{+0.05}_{-0.05}$ & $0.28^{+0.57}_{-0.18}$ & $1.88^{+0.18}_{-0.14}$ & $14.44^{+0.85}_{-0.68}$ \\ 
BP Tau    & $3703^{+48}_{-35}$ & $4.26^{+0.07}_{-0.07}$ & $1.36^{+0.07}_{-0.06}$ & $2.03^{+0.22}_{-0.03}$ & $2.38^{+0.15}_{-0.14}$ & $9.94^{+2.61}_{-0.99}$ \\ 
CI Tau    & $3888^{+57}_{-73}$ & $3.74^{+0.10}_{-0.14}$ & $1.87^{+0.08}_{-0.08}$ & $1.19^{+0.69}_{-0.69}$ & $1.78^{+0.11}_{-0.12}$ & $12.71^{+0.40}_{-0.48}$ \\ 
CX Tau    & $3498^{+47}_{-53}$ & $3.64^{+0.10}_{-0.12}$ & $0.26^{+0.03}_{-0.02}$ & $2.32^{+0.43}_{-0.44}$ & $ 1.10^{+0.16}_{-0.16}$ & $21.41^{+0.32}_{-0.34}$ \\ 
CY Tau    & $3403^{+43}_{-21}$ & $3.82^{+0.06}_{-0.06}$ & $0.58^{+0.04}_{-0.04}$ & $1.84^{+0.56}_{-0.44}$ & $1.39^{+0.08}_{-0.09}$ & $10.55^{+0.54}_{-0.33}$ \\ 
DE Tau    & $3459^{+40}_{-52}$ & $3.67^{+0.07}_{-0.08}$ & $1.15^{+0.04}_{-0.04}$ & $2.49^{+0.34}_{-0.33}$ & $0.75^{+0.10}_{-0.10}$ & $10.33^{+0.33}_{-0.40}$ \\ 
DK Tau A  & $3501^{+43}_{-29}$ & $3.71^{+0.08}_{-0.09}$ & $1.90^{+0.08}_{-0.08}$ & $0.98^{+0.53}_{-0.56}$ & $2.33^{+0.16}_{-0.16}$ & $17.40^{+0.88}_{-0.72}$ \\ 
DK Tau B  & $3411^{+67}_{-18}$ & $3.75^{+0.06}_{-0.06}$ & $1.10^{+0.06}_{-0.04}$ & $1.58^{+0.37}_{-0.40}$ & $0.81^{+0.12}_{-0.12}$ & $9.47^{+0.34}_{-0.68}$ \\ 
DL Tau    & $3915^{+39}_{-41}$ & $3.81^{+0.07}_{-0.07}$ & $2.97^{+0.07}_{-0.08}$ & $0.40^{+0.59}_{-0.30}$ & $2.05^{+0.07}_{-0.09}$ & $9.77^{+0.66}_{-0.37}$ \\ 
DM Tau    & $3302^{+23}_{-14}$ & $3.90^{+0.02}_{-0.03}$ & $0.03^{+0.02}_{-0.01}$ & $0.14^{+0.20}_{-0.04}$ & $1.71^{+0.06}_{-0.08}$ & $5.15^{+0.07}_{-1.28}$ \\ 
DN Tau    & $3520^{+44}_{-27}$ & $3.65^{+0.07}_{-0.06}$ & $0.62^{+0.02}_{-0.02}$ & $0.90^{+0.34}_{-0.38}$ & $1.59^{+0.05}_{-0.06}$ & $11.01^{+0.21}_{-0.21}$ \\
DoAr 25   & $3696^{+58}_{-56}$ & $3.52^{+0.13}_{-0.10}$ & $0.50^{+0.03}_{-0.03}$ & $0.38^{+0.56}_{-0.28}$ & $2.09^{+0.13}_{-0.12}$ & $15.90^{+0.60}_{-0.34}$\\ 
DoAr 33   & $3697^{+22}_{-35}$ & $3.86^{+0.04}_{-0.05}$ & $0.70^{+0.01}_{-0.02}$ & $0.16^{+0.25}_{-0.06}$ & $2.16^{+0.07}_{-0.06}$ & $12.68^{+0.35}_{-0.33}$\\ 
DoAr 43 SW & $4221^{+67}_{-56}$ & $3.65^{+0.13}_{-0.20}$ & $1.96^{+0.14}_{-0.17}$ & $0.19^{+0.43}_{-0.09}$ & $2.04^{+0.39}_{-0.47}$ & $34.57^{+1.13}_{-2.06}$\\ 
DS Tau    & $3744^{+66}_{-67}$ & $3.81^{+0.10}_{-0.11}$ & $1.61^{+0.05}_{-0.05}$ & $0.90^{+0.47}_{-0.67}$ & $2.23^{+0.13}_{-0.11}$ & $14.07^{+0.43}_{-0.42}$ \\ 
FM Tau    & $3378^{+68}_{-88}$ & $3.88^{+0.06}_{-0.08}$ & $1.86^{+0.11}_{-0.11}$ & $0.28^{+0.73}_{-0.18}$ & $2.86^{+0.18}_{-0.18}$ & $10.25^{+0.74}_{-1.20}$ \\ 
FP Tau    & $3395^{+82}_{-63}$ & $3.40^{+0.20}_{-0.12}$ & $0.26^{+0.03}_{-0.02}$ & $1.81^{+0.71}_{-0.51}$ & $1.14^{+0.25}_{-0.21}$ & $33.47^{+0.65}_{-0.62}$  \\ 
FX Tau A  & $3552^{+51}_{-53}$ & $3.81^{+0.07}_{-0.08}$ & $0.69^{+0.04}_{-0.03}$ & $1.78^{+0.45}_{-0.43}$ & $0.99^{+0.07}_{-0.07}$ & $5.62^{+0.34}_{-0.19}$  \\ 
GK Tau A  & $3655^{+57}_{-59}$ & $3.61^{+0.12}_{-0.11}$ & $1.74^{+0.05}_{-0.05}$ & $0.90^{+0.41}_{-0.42}$ & $2.23^{+0.13}_{-0.13}$ & $20.35^{+0.55}_{-0.54}$ \\ 
GM Aur    & $3843^{+43}_{-57}$ & $3.91^{+0.06}_{-0.08}$ & $0.59^{+0.02}_{-0.03}$ & $0.17^{+0.31}_{-0.07}$ & $1.99^{+0.10}_{-0.12}$ & $14.06^{+0.34}_{-0.52}$ \\ 
GO Tau    & $3429^{+48}_{-31}$ & $3.86^{+0.05}_{-0.05}$ & $0.28^{+0.03}_{-0.02}$ & $0.93^{+0.34}_{-0.49}$ & $1.48^{+0.09}_{-0.11}$ & $13.44^{+0.48}_{-0.28}$ \\ 
GSS 26    & $3498^{+87}_{-83}$ & $3.44^{+0.24}_{-0.10}$ & $0.53^{+0.03}_{-0.04}$ & $0.78^{+0.56}_{-0.67}$ & $2.57^{+0.16}_{-0.17}$ & $17.94^{+0.53}_{-0.71}$\\ 
GSS 37 W  & $3590^{+31}_{-47}$ & $3.45^{+0.12}_{-0.06}$ & $0.47^{+0.03}_{-0.02}$ & $0.91^{+0.44}_{-0.44}$ & $1.87^{+0.10}_{-0.10}$ & $16.73^{+0.43}_{-0.45}$\\ 
GSS 39    & $3396^{+63}_{-83}$ & $3.20^{+0.10}_{-0.16}$ & $1.82^{+0.07}_{-0.07}$ & $1.57^{+0.47}_{-0.43}$ & $1.85^{+0.18}_{-0.19}$ & $26.28^{+0.75}_{-0.63}$ \\
Haro 1-16 & $3953^{+109}_{-118}$ & $3.93^{+0.09}_{-0.14}$ & $2.33^{+0.15}_{-0.17}$ & $0.31^{+0.77}_{-0.21}$ & $2.47^{+0.26}_{-0.27}$ & $16.17^{+1.14}_{-1.01}$\\ 
Haro 6-13 & $3700^{+96}_{-109}$ & $3.54^{+0.21}_{-0.25}$ & $1.55^{+0.12}_{-0.13}$ & $1.96^{+1.02}_{-1.02}$ & $1.15^{+0.28}_{-0.28}$ & $22.20^{+0.83}_{-0.80}$\\ 
Haro 6-33 & $3622^{+62}_{-51}$ & $3.69^{+0.12}_{-0.14}$ & $0.55^{+0.03}_{-0.03}$ & $1.52^{+0.67}_{-0.56}$ & $1.35^{+0.19}_{-0.24}$ & $23.70^{+0.63}_{-0.36}$\\ 
HK Tau  A & $3509^{+48}_{-46}$ & $3.62^{+0.09}_{-0.10}$ & $0.74^{+0.03}_{-0.03}$ & $2.48^{+0.47}_{-0.41}$ & $1.15^{+0.14}_{-0.14}$ & $22.39^{+0.47}_{-0.26}$ \\ 
HO Tau    & $3427^{+50}_{-45}$ & $4.34^{+0.07}_{-0.09}$ & $1.08^{+0.05}_{-0.05}$ & $2.54^{+0.90}_{-0.53}$ & $0.93^{+0.30}_{-0.29}$ & $22.66^{+0.72}_{-0.61}$  \\ 
HP Tau    & $4119^{+39}_{-39}$ & $3.81^{+0.09}_{-0.11}$ & $1.80^{+0.09}_{-0.08}$ & $0.46^{+0.68}_{-0.36}$ & $2.13^{+0.12}_{-0.14}$ & $17.27^{+0.70}_{-0.51}$ \\ 
IP Tau    & $3659^{+37}_{-41}$ & $4.34^{+0.05}_{-0.05}$ & $1.24^{+0.05}_{-0.04}$ & $2.03^{+0.17}_{-0.03}$ & $2.64^{+0.11}_{-0.11}$ & $12.67^{+0.39}_{-1.23}$\\ 
IQ Tau    & $3549^{+60}_{-54}$ & $3.60^{+0.12}_{-0.13}$ & $0.91^{+0.04}_{-0.04}$ & $1.72^{+0.53}_{-0.47}$ & $1.85^{+0.11}_{-0.11}$ & $15.59^{+0.33}_{-0.39}$\\ 
LkCa 15   & $4093^{+41}_{-45}$ & $4.16^{+0.08}_{-0.19}$ & $1.07^{+0.05}_{-0.06}$ & $2.04^{+0.36}_{-0.49}$ & $1.97^{+0.13}_{-0.13}$ & $15.15^{+0.49}_{-0.89}$\\ 
ROX 25    & $4400^{+46}_{-51}$ & $3.95^{+0.22}_{-0.13}$ & $1.05^{+0.19}_{-0.15}$ & $2.56^{+1.10}_{-1.50}$ & $0.71^{+0.57}_{-0.61}$ & $28.81^{+0.60}_{-0.57}$\\ 
ROX 27    & $3760^{+51}_{-58}$ & $3.60^{+0.11}_{-0.13}$ & $1.28^{+0.04}_{-0.05}$ & $0.94^{+0.49}_{-0.55}$ & $1.83^{+0.11}_{-0.13}$ & $18.21^{+0.44}_{-0.41}$\\ 
UY Aur NE & $3603^{+45}_{-64}$ & $3.42^{+0.08}_{-0.10}$ & $1.23^{+0.05}_{-0.05}$ & $1.06^{+0.58}_{-0.62}$ & $1.75^{+0.12}_{-0.13}$ & $21.03^{+0.40}_{-0.48}$\\ 
V710 Tau N & $3484^{+20}_{-35}$ & $3.76^{+0.04}_{-0.05}$ & $0.27^{+0.02}_{-0.02}$ & $1.99^{+0.45}_{-0.41}$ & $1.83^{+0.09}_{-0.09}$ & $17.96^{+0.34}_{-0.44}$\\ 
V710 Tau S & $3400^{+10}_{-7}$ & $3.78^{+0.03}_{-0.03}$ & $0.22^{+0.01}_{-0.02}$ & $1.67^{+0.27}_{-0.31}$ & $0.88^{+0.17}_{-0.14}$ & $25.12^{+0.30}_{-0.22}$\\ 
WSB 82 & $3968^{+142}_{-169}$ & $3.99^{+0.22}_{-0.44}$ & $4.15^{+0.44}_{-0.70}$ & $3.64^{+0.36}_{-2.08}$ & $3.24^{+0.63}_{-0.75}$ & $32.03^{+0.83}_{-0.96}$\\ 
YLW 58   & $3004^{+110}_{-37}$ & $ 3.10^{+0.37}_{-0.10}$ & $1.68^{+0.11}_{-0.15}$ & $1.06^{+1.00}_{-0.70}$ & $1.18^{+0.15}_{-0.20}$ & $12.14^{+1.00}_{-1.20}$ \\ 
\enddata
\tablecomments{The reported uncertainties correspond to 3$\sigma$ deviations from the median value obtained from the MCMC distributions.}
\end{deluxetable*}

%% file: Stellar_masses.tex
\begin{deluxetable*}{lcccccc}
\tabletypesize{\scriptsize}
\tablecaption{Optical parameters from literature and derived masses. \label{table:Stellar_masses}}
\tablehead{
\colhead{Source} & \colhead{Adopted} & \colhead{$\rm T_{optical}$} & \colhead{Reference} &\colhead{$\rm M_{\rm K-band}$} & \colhead{$\rm M_{\rm optical}$} & \colhead{$\rm M_{\rm dyn}$} \\
\colhead{name}& \colhead{Sp Type} & \colhead{(K)} & \colhead{} & \colhead{($\rm M_\odot$)} & \colhead{($\rm M_\odot$)} & \colhead{($\rm M_\odot$)} }
\startdata
AA Tau & M0.6 & $3792\pm36$ & (1) & $0.66_{-0.07}^{+0.07} $ & $0.90_{-0.05}^{+0.06} $ & $0.84_{-0.04}^{+0.04} $  \\ 
BP Tau & M0.5 & $3810\pm36$ & (1) & $0.73_{-0.05}^{+0.05} $ & $0.83_{-0.05}^{+0.04} $ & $1.10_{-0.04}^{+0.04} $ \\ 
CI Tau & K5.5 & $4020\pm48$ & (1) & $1.03_{-0.08}^{+0.07} $ & $1.30_{-0.05}^{+0.04} $ & $0.90_{-0.02}^{+0.02} $ \\ 
CX Tau & M2.5 & $3485\pm30$ & (1) & $0.59_{-0.07}^{+0.07} $ & $0.57_{-0.05}^{+0.06} $ & $0.38_{-0.02}^{+0.02} $ \\ 
CY Tau & M2.3 & $3515\pm30$ & (1) & $0.45_{-0.05}^{+0.05} $ & $0.57_{-0.04}^{+0.04} $ & $0.30_{-0.02}^{+0.02} $ \\ 
DE Tau & M2.3 & $3515\pm30$ & (1) & $0.55_{-0.07}^{+0.06} $ & $0.60_{-0.04}^{+0.05} $ & $0.41_{-0.03}^{+0.03} $ \\ 
DK Tau A & K8.5 & $3930\pm30$ & (1) & $0.58_{-0.06}^{+0.07} $ & $1.08_{-0.04}^{+0.04} $ & $0.55_{-0.13}^{+0.13} $ \\ 
DK Tau B & M1.7 & $3608\pm32$ & (1) & $0.47_{-0.05}^{+0.05} $ & $0.68_{-0.04}^{+0.05} $ & $1.19_{-0.27}^{+0.27} $ \\ 
DL Tau & K5.5 & $4163\pm48$ & (1) & $1.05_{-0.06}^{+0.05} $ & $1.29_{-0.05}^{+0.05} $ & $1.04_{-0.02}^{+0.02} $ \\ 
DM Tau & M3 & $3410\pm44$ & (1) & $0.35_{-0.04}^{+0.05} $ & $0.45_{-0.04}^{+0.04} $ & $0.55_{-0.02}^{+0.02} $ \\ 
DN Tau & M0.3 & $3846\pm36$ & (1) & $0.63_{-0.08}^{+0.06} $ & $1.01_{-0.05}^{+0.05} $ & $0.87_{-0.15}^{+0.15} $ \\ 
DoAr 25 & K5 & $4210\pm262$  & (3) & $0.89_{-0.09}^{+0.09} $ & $1.38_{-0.23}^{+0.13} $ & \\ 
DoAr 33 & K5.5 & $4163\pm130$  & (3) & $0.76_{-0.06}^{+0.07} $ & $1.28_{-0.13}^{+0.10} $ & \\ 
DoAr 43 SW & K2 & $4710\pm166$  & (6) & $1.39_{-0.10}^{+0.02}$  &  $1.61_{-0.08}^{>0.09}$  & \\ 
DS Tau & M0.4 &  $3828\pm36$ & (1) & $0.83_{-0.08}^{+0.10} $ & $0.95_{-0.06}^{+0.05} $ & $0.83_{-0.02}^{+0.02} $ \\ 
FM Tau & M0 & $3900\pm72$ & (2) & $0.42_{-0.07}^{+0.08} $ & $1.01_{-0.09}^{+0.08} $ & $0.36_{-0.02}^{+0.02} $ \\ 
FP Tau & M2.6 & $3470\pm30$ & (1) & $0.56_{-0.11}^{+0.09} $ & $0.65_{-0.16}^{+0.04} $ & $0.36_{-0.02}^{+0.02} $\\ 
FX Tau A & M2.2 & $3530\pm30$ & (1) & $0.61_{-0.06}^{+0.07} $ & $0.59_{-0.04}^{+0.04} $ & \\ 
GK Tau A & K6.5 & $4068\pm48$ & (1) & $0.79_{-0.09}^{+0.09} $ & $1.24_{-0.06}^{+0.05} $ & $0.73_{-0.06}^{+0.06} $ \\ 
GM Aur & K6 & $4115\pm48$ & (1) & $0.94_{-0.07}^{+0.06} $ & $1.22_{-0.05}^{+0.05} $ & $1.14_{-0.02}^{+0.02} $ \\ 
GO Tau & M2.3 & $3900\pm72$ & (2) & $0.47_{-0.05}^{+0.05} $ & $1.02_{-0.09}^{+0.08} $ & $0.45_{-0.01}^{+0.01} $ \\
GSS 26 &  &   &  & $0.68_{-0.18}^{+0.11} $ &  & \\ 
GSS 37 W & M1 & $3720\pm170$ & (4) & $0.79_{-0.10}^{+0.06} $ & $0.93_{-0.24}^{+0.16} $ & \\
GSS 39 &  &  &  & $0.40_{-0.01}^{+0.18} $ &  & $0.47_{-0.04}^{+0.04} $ \\ 
Haro 1-16 & K3 & $4543\pm290$ & (7) & $1.07_{-0.16}^{+0.13}$ & $1.49_{-0.21}^{>0.21}$  & \\ 
Haro 6-13 & M0.5 & $3810\pm90$ & (5) & $0.88_{-0.14}^{+0.12} $ & $1.00_{-0.13}^{+0.10} $ & $0.93_{-0.14}^{+0.14}$  \\ 
Haro 6-33 & M1.6 & $3620\pm36$ & (1) & $0.72_{-0.07}^{+0.09} $ & $0.72_{-0.05}^{+0.07} $ & $0.60_{-0.10}^{+0.10}$ \\ 
HK Tau A & M1.5 & $3640\pm32$ & (1) & $0.61_{-0.06}^{+0.08} $ & $0.77_{-0.06}^{+0.06} $ & $0.53_{-0.03}^{+0.03}$ \\ 
HO Tau & M3.2 & $3810\pm72$ & (2) & $0.43_{-0.05}^{+0.06} $ & $0.80_{-0.07}^{+0.07} $ & $0.43_{-0.03}^{+0.03}$ \\ 
HP Tau & K4 & $4377\pm60$ & (1) & $1.25_{-0.05}^{+0.05} $ & $1.43_{-0.04}^{+0.03} $ & \\ 
IP Tau & M0.6 & $3792\pm36$ & (1) & $0.68_{-0.05}^{+0.05} $ & $0.79_{-0.06}^{+0.06} $ & $0.94_{-0.05}^{+0.05}$ \\ 
IQ Tau & M1.1 & $3704\pm32$  & (1) & $0.67_{-0.09}^{+0.09} $ & $0.86_{-0.06}^{+0.06} $ & $0.74_{-0.02}^{+0.02}$ \\ 
LkCa 15 & K5.5 & $4163\pm48$  & (1) & $1.09_{-0.12}^{+0.08} $ & $1.13_{-0.14}^{+0.09} $ & $1.14_{-0.03}^{+0.03}$ \\ 
ROX 25 & K6 & $4115\pm95$  & (4) & $1.41_{-0.11}^{+0.05} $ & $1.21_{-0.10}^{+0.10} $ & $1.27_{-0.07}^{+0.07}$ \\ 
ROX 27 & K5 & $4210\pm262$  & (3) & $0.93_{-0.08}^{+0.08} $ & $1.36_{-0.24}^{+0.14} $ & \\ 
UY Aur NE & M0 & $3900\pm90$  & (8) & $0.80_{-0.12}^{+0.06} $ &  & \\ 
V710 Tau N & M1.7* & $3608\pm44$  & (1) & $0.55_{-0.06}^{+0.05} $ & $0.68_{-0.05}^{+0.07} $ & $0.67_{-0.06}^{+0.06}$ \\
V710 Tau S & M3.3* & $3344\pm32$  & (1) & $0.45_{-0.05}^{+0.05} $ & $0.40_{-0.03}^{+0.03} $ & \\ 
WSB 82 &  &  & & $1.06_{-0.20}^{+0.16}$ &  & \\ 
YLW 58 & M4.5 & $3085\pm210$  & (3) &  &  & $0.10_{-0.01}^{+0.01}$ \\ 
\enddata
\tablecomments{(1) is \cite{Herczeg2014}, (2) is \cite{Luhman2010}}, (3) is \cite{Wilking2005}, (4) is \cite{Erickson2011}, (5) is \cite{White2004}, (6) is \cite{Torres2006}, (7) is \cite{Bouvier1992}, and (8) is \cite{Hartigan2003}.\\
\tablecomments{As discussed in \cite{Manara2019}, \cite{Herczeg2014} defined V710 Tau A as the southern companion and V710 Tau B as the Northern companion.}
\end{deluxetable*}